\documentclass[twocolumn,aps,pra,longbibliography]{revtex4-2}
\usepackage{amsmath}
\usepackage{amssymb}
\usepackage{graphicx}
\usepackage[utf8]{inputenc}
\usepackage[T1]{fontenc}
\usepackage{color}
\newcommand{\av }[1]{\langle#1\rangle}
\renewcommand{\i}{\mathrm{i}}
\usepackage{braket}

\begin{document}

\title{Correlated volumes for extended wavefunctions on a random-regular graph}

\author{Manuel Pino}
\address{
 Nanotechnology Group, USAL-Nanolab,
Universidad de Salamanca, E-37008 Salamanca, Spain.\\
 Institute of Fundamental Physics and Mathematics, E-37008 Salamanca\\
 Institute of Fundamental Physics IFF-CSIC, Calle Serrano 113b, Madrid 28006, Spain
}

\author{Jose E. Roman}
\address{%
 DSIC, Universitat Polit\`ecnica de Val\`encia, Cam\'{\i} de Vera s/n, 46022 Val\`encia, Spain}%

\begin{abstract}
{  We study the metallic phase of the Anderson model in a random-regular graph, specifically the degree of ergodicity of the high-energy wavefunctions.  We use the multifractal formalism to analyze numerical data for unprecedented large system sizes, obtaining a set of correlated volumes $N_q$ which control finite-size effects of wavefunction $q$-moment. Those volumes grow very fast $\log(\log(N_q))\sim W,$ with disorder strength but show no tendency to diverge, at least in an intermediate metallic regime. Close to the Anderson transitions, we characterize the crossover to system sizes much smaller than the first correlated volume. Once this crossover has taken place, we obtain evidence of a scaling in which the derivative of the first fractal dimension behaves critically with an exponent $\nu=1.$}
\end{abstract}
\maketitle

\section{Introduction}

Non-ergodic and chaotic quantum states exist on some single-particle models with infinite range connectivity\ \cite{aizenman2011extended,warzel2013resonant,Kravtsov2015, khaymovich2021dynamical, pino2019ergodic}. However, its role in realistic many-body systems---those in which interactions have a finite range---has been the subject of intense research in recent years\ \cite{Al1997,Pino15,Kravtsov2015,venturelli2022replica, Tikhonov2016b,tikhonov2019}. To clarify this situation, we can study systems that interpolate between infinite and finite connectivity. That is the case of a particle hopping in a random-regular graph (RRG) in which, although having finite connectivity,  the number of closed loops is small due to the hierarchical structure of the lattice\ \cite{dorogovtsev2008critical, mezard2001bethe}. Notice that the effect of closed loops is neglected in mean-field theories, the ones that usually explain infinite-dimensional models. Besides its relative simplicity, the hopping process of a particle in a RRG may capture some key ingredients of the effective hopping in the many-body space of interacting Hamiltonians\ \cite{Al1997,Basko2006}. Furthermore, non-ergodic and chaotic quantum systems are likely to play an important role in the field of quantum computing\ \cite{altshuler2010anderson,laumann2015quantum,smelyanskiy2020nonergodic, faoro2018non, Pino_Ripoll_2020,pino2013capturing}.

The metal-insulator transition in RRGs has recently been the subject of deep analysis. The first works on the subject point to a non-ergodic phase, at least for the metal near the Anderson transition\ \cite{Deluca2014,Cuevas2001, kravtsov2018non}, which is consistent with its reported slow dynamics\ \cite{biroli2020anomalous,colmenarez2022}. However, several other works argued that non-ergodicity is a finite-size effect\ \cite{Tikhonov2016b, Tikhonov2016, tikhonov2019,garcia2017scaling,garc2019,biroli2020anomalous,biroli2018delocalization,Biroli2022},  which disappear for sizes $N$ larger than a typical volume $N_e,$ referred to as ergodic volume. Following those references, non-ergodicity can still play an important role near the critical disorder $W_c$ as $N_e$ diverges very fast $\log(N_e)\sim (W-W_c)^{-1/2}$ upon approaching the Anderson transition\ \cite{mirlin1994distribution,Mirlin_Fyodorov_1991, Tikhonov_Mirlin_2021}. Arguments for the ergodicity of the metal are mainly based on the mean-field solutions for the imaginary part of the local Green's function\ \cite{tikhonov2018,Tikhonov2016, Tikhonov2016b}, which has been obtained with the supersymmetric formalism\ \cite{efetov1985anderson, Zirnbauer1986} and with population-dynamics-like algorithms\ \cite{biroli2018delocalization, Biroli2022, Biroli2017, garcia2017scaling}.  Refs.\ \cite{Biroli2022,Biroli2017,garcia2017scaling,garc2019,Tikhonov_Mirlin_2021} also report evidence of a divergence in the ergodic volume with exponent $\nu=1/2$ obtained from exact diagonalization.

Although exact diagonalization has considerably helped to understand Anderson transitions\ \cite{Edwards1972, rodriguez2010critical,ortuno2009random}, those techniques have not yet given conclusive results regarding the ergodicity of metallic wavefunctions in RRGs. This is because finite size effects are very slow, they are controlled by the logarithm of system size, together with the need for eigenstate extraction at the middle of the spectrum, where Lanczos methods work poorly. Here, we have partially overcome this last limitation using a new polynomial filter implemented in the SLEPC libraries~\cite{Hernandez:2005:SSF} to reach much larger sizes than previously obtained, up to $N=4\times 10^6$ sites~\cite{Roman2024_preparation}. To deal with the slowness of the finite-size effect, we have developed an accurate procedure to extract the properties of wavefunctions, based on the most general form of the corrections within the multifractal ansatz.

Using the methods described in the previous paragraph, we obtain $D_q$ and their corresponding correlated volumes $N_q$ from the moments of the wavefunction $I_q=N\av{|\psi^{2q}_i|}$ which are related via $I_q= \left(N/N_q\right)^{(1-q)D_q}.$ Our results are compatible with an ergodic metal $D_q=1$ and correlated volumes given by $\log(\log(N_q))\approx  W/4+A_q$ for disorder in $5<W<15.$  The existence of many correlated volumes that are { roughly} related by $N_q=(N_{q^\prime})^{c_{qq^\prime}}$ implies that there is not a single ergodic volume, but a bunch of correlated volumes $N_q$ that characterize finite-size effects. Alternatively,  there is a single graph diameter $\log(N_q)$ that sets a length scale at which finite-size effects become small.  We do not find numerical evidence of diverging correlated volumes for disorders smaller than  $W\approx 15 $ in contrast to the reported numerical analysis in Refs.\ \cite{biroli2018delocalization,Biroli2022,Tikhonov_Mirlin_2021,garcia2017scaling}. We found that one of the correlated volumes $N_0$ is close to the ergodic volume $N_e$ obtained from Refs.\ \cite{biroli2018delocalization, Tikhonov_Mirlin_2021}  but {  slightly} smaller in the region $W=[13,15]$. We cannot extract from our fittings procedure any correlated volume closer to the Anderson transition $15.5<W<W_c\approx 18.1.$ This may be due to correlated volumes becoming much larger than system sizes. But it could also be caused by the existence of a genuine non-ergodic metal in this region, as in our procedure we fix ergodicity to produce fittings with less free parameters. Finally, we have found evidence of critical behaviour in the first fractal dimension after the crossover $N\ll N_1$ took place, similar to Ref.\ \cite{pino2020scaling}.

\section{Model and multifractal ansatz}\label{sec:model}

We study the Anderson model\ \cite{An1958, Ab1979} of a particle hopping between the $N$ nodes of a RRG. That is, we generate a Hamiltonian
\begin{equation}
 H= \sum_{i=1}^{N} \phi_i c_i^\dagger c_i + \sum_{\av{ij}} (c_i^\dagger c_j+c_j^\dagger c_i),\label{Eq:model}
\end{equation}
where $ c_{i},\ c_i^\dagger $ are destruction and creation operators at site $i $ and $\phi_i$ are random numbers in $[-\frac{W}{2},\frac{W}{2}],$ being $W$ the disorder. The sum at the right-hand side of Eq.~\ref{Eq:model} runs for all the edges $\av{ij}$ of a graph which is generated following the probability distribution of a RRG  with branching number $k=2$\ \cite{pythonnetwokx}. Several previous references agreed that the Anderson transition for this model is at $W_c\approx18.17$\ \cite{tikhonov2019,pino2020scaling,sierant2023universality}.

\begin{figure}[t!]
\includegraphics[width=1\columnwidth]{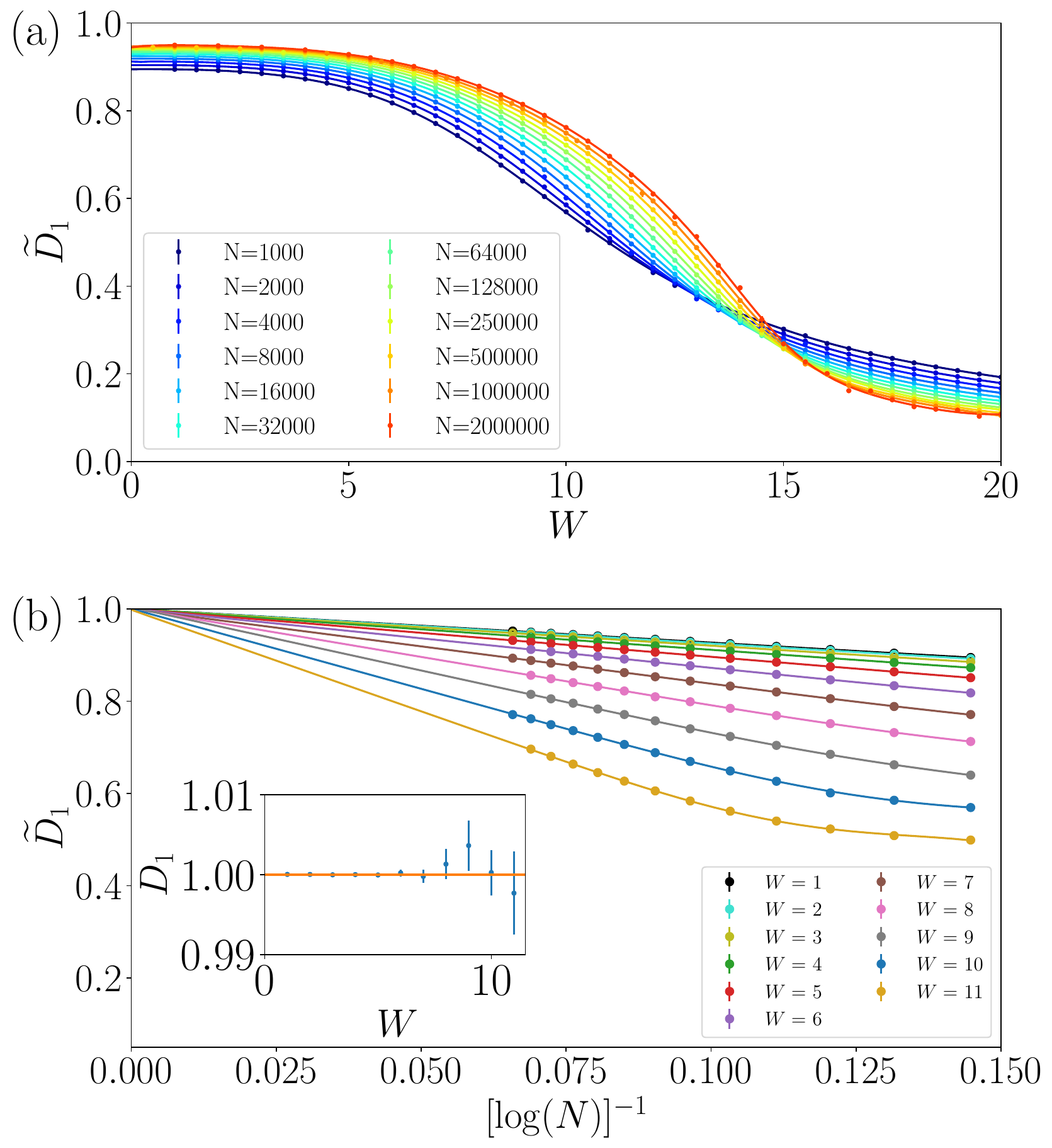}
\caption{\label{fig:D1}
(a) Effective first fractal dimension $\widetilde{D}_1 =S/\log(N),$ with $S$ participation entropy, as a function of disorder strength $W$ for several system sizes N. (b) Effective fractal dimension as a function of system sizes for several values of the disorder between $W=1$ and $W=11.$ The solid lines fit all data to $\sum_{j=0,k=0}^{r,s} a_{jk}(W)/N^j[\log(N)]^k$ with $r=s=1$ for all available sizes at each disorder value $N=10^3\dots 2\times 10^6$ ($N=4\times 10^6$ for $W=1,3,5,7,10$). All the fittings have $p$-values larger than $0.1,$ except $W=2,6$ which is quite close $p>0.07.$ Fractal dimension at the thermodynamic limit can be obtained as $D_1 = a_{00}$ (inset).
}
\end{figure}

{  The eigenstates of Hamiltonian in Eq.~\ref{Eq:model} are analyzed via the multifractal formalism\ \cite{Kadanoff1986}. We denote the amplitude of eigenstate $\psi$ at site $i$ as $\psi_i$ and assume that the support set of sites that scale as $\alpha=-\log_N(|\psi_i|^2)$  is given by $N^{f(\alpha)},$ with $f(\alpha)$ the multifractal spectrum. The Legendre transforms of $f$ are the multifractal dimensions $D_q=[f(\alpha_q)-\alpha_qq]/(q-1)$ with $f^\prime(\alpha_q)=q$. An ergodic system implies that $D_q=1$, while $D_q<1$ occurs for a non-ergodic wavefunction. We extract $D_q$ averaging  the closest to zero wavefunction moment $I_q=\sum_{i=1}^N|\psi_i|^{2q}$ over Hamiltonian realizations\ \cite{rodriguez2010critical, rodriguez2011}. We postulate a scaling form of the effective fractal dimensions $\widetilde{D}_q = \log_N{I_q}/(1-q)$ given by
\begin{align}
\widetilde{D}_q = \sum_{\substack{j=0,\dots,r\\ k=0,\dots,s}} \frac{a^{(q)}_{jk}}{N^j[\log(N)]^k},\label{eq:corr}
\end{align}
with $r,s$ integers. The quantities $a^{(q)}_{jk}$ provide information about the multifractal nature of the wavefunctions and their finite-size corrections. For instance,  fractal dimensions correspond to $D_q=a^{(q)}_{00}$ while we define the $q$-correlated volume as the exponential of the leading correction $\log(N_q)=a^{(q)}_{01},$  such that we can express  $\widetilde{D}_q=D_q\left[1-\log_N(N_q)\right]$ at the leading order.

We will fit our numerical data to Eq.\ \ref{eq:corr} with $r=s=1$\  \cite{pausch2021chaos} and accept the result if the goodness of the fit is acceptable (see Appendix\ \ref{sec:app2} for an analysis with $r,s>1$). The form Eq.\ \ref{eq:corr} will also be used for the exponential of the typical value of the wavefunction $\widetilde{\alpha}_0=-\av{\log_N(|\psi_i|^2)}.$ Notice that this quantity is related to $I_q$ in virtue of $\av{\log(|\psi_i|^2)}=(I_q/N-1)/q$  in the limit $q\rightarrow 0.$ We notice that leading finite-size corrections in $\widetilde{D}_q$ become negligible when $\log(N)\gg \log(N_q),$ so that $\log(N_q)$ marks the diameter at which $\widetilde{D}_q$ is close to its thermodynamics limit value. Those quantities play a similar role as the one attributed to the so-called ``ergodic'' volume $N_e=-\log({\rm<Im G>_{typ}})$ in Refs.\ \cite{Tikhonov_Mirlin_2021,garc2019, garcia2017scaling, biroli2018delocalization}, see Appendix\ \ref{sec:app2}. }

\section{Fractal dimensions and correlated volumes in the metal}\label{sec:corr_v}

We extract $D_1$ and $\log(N_1)$ from the data for $\widetilde{D}_1.$ Effective fractal dimension $\widetilde{D}_1$ appears in Fig.\ \ref{fig:D1} as a function of disorder $W$ for sizes $N=10^3$ to $N=2\times 10^6,$ together with Padé approximants (solid lines) for each size. The ergodic limit $D_1=1,$ which there is no doubt occurs deep enough in the metal, is not reached even for the largest system size at small disorder. However, we can reliably extract fractal dimensions by fitting $\widetilde{D}_1$ for all available sizes via  Eq.~\ref{eq:corr} with $r=s=1,$ panel (b) of Fig.\ \ref{fig:D1}. The result is compatible with ergodicity $D_1=1$ up to the largest disorder computed, inset of panel (b) of Fig.\ \ref{fig:D1}.  We notice that fitting without the $1/N$ correction produces very bad quality fitting even for small disorders. We have only fitted up to disorder $W=11$ as the fitting for larger disorder produces too small $p$-values, all $p$-values are reported in Fig.\ \ref{Fig:D1_details} of Appendix\ \ref{supp:fittings}. This implies that larger $r,s$ should be used in order to correctly describe \emph{all} available data sizes for $W\geq 12.$

\begin{figure}[t!]
\includegraphics[width=0.99\columnwidth]{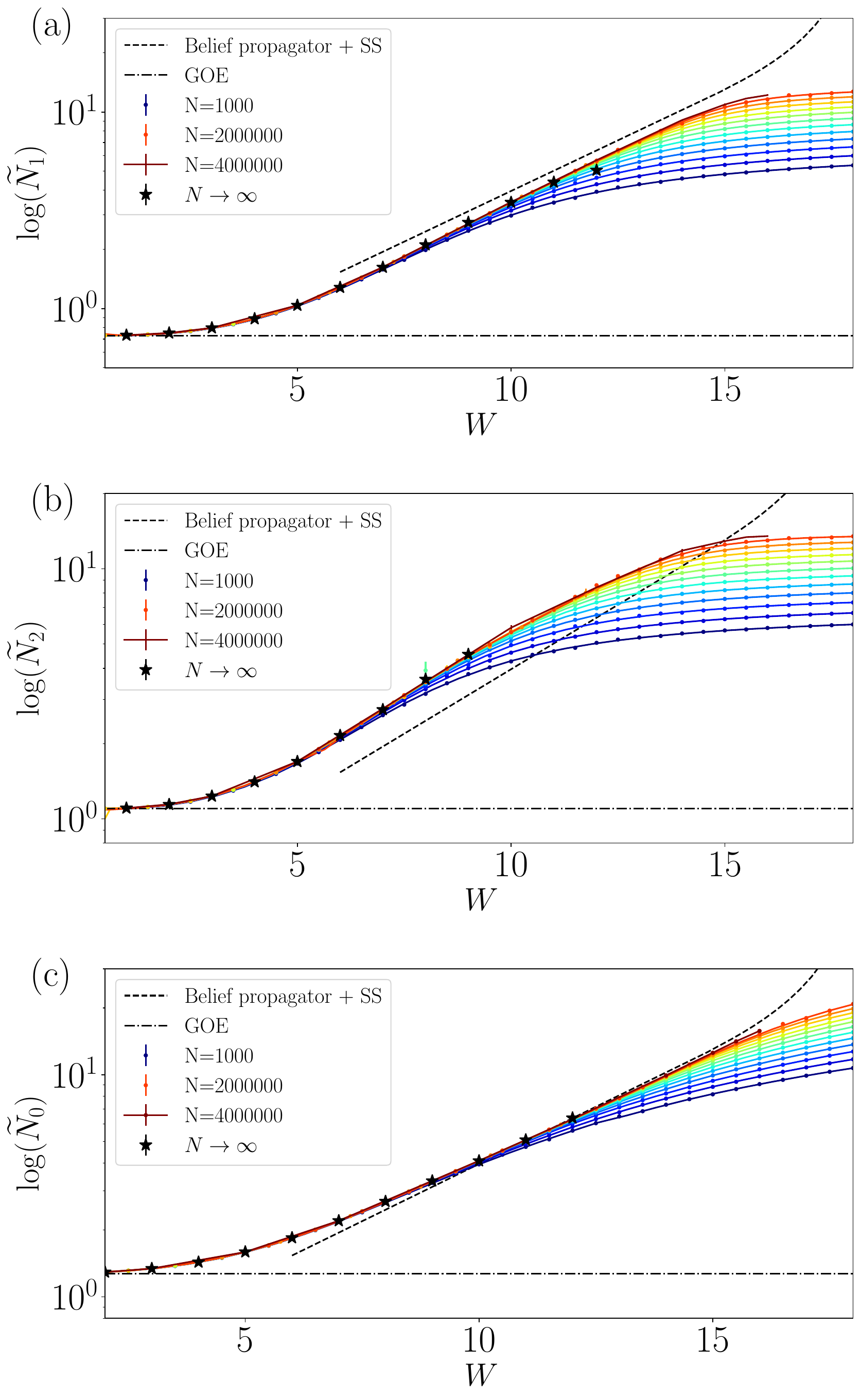}
\caption{\label{fig:loglo_x1}
Finite-size corrections  $\log(\widetilde{N}_q)$ assuming ergodicity for $q=1$ (a) $q=2$ (b) and $q=M$ (c) as a function of disorder $W$ for several system sizes $N.$ {  Notice the log scale in y-axe.} Solid lines are Padé approximants for sizes $N=10^3,\dots,2\times 10^6$ while for $N=4\times 10^6$ are just guide to the eyes. Stars are the value of $\log(\widetilde{N}_q)$ in the thermodynamic limit as the coefficient $\log(N_q) = a^{(q)}_{01}$ when fitting $\widetilde{D}_q=\sum_{j=0,k=0}^{1} a^{(q)}_{jk}(W)/(N^j[\log(N)]^k)$ for all the available sizes $N=10^3,\dots,4\times 10^6$ at each disorder value. We do not assume ergodicity for those fitting but we obtain ergodic behaviour $D_q =a_{00}=1$ up to fitting uncertainties. The dashed line is the typical imaginary part of the self-energy obtained via belief propagation and super-symmetric formalism in Refs.\ \cite{biroli2018delocalization, Tikhonov_Mirlin_2021} and the semi-dashed line is the GOE limit.
}
\end{figure}

We focus on the determination of the correlated volumes now. They can be extracted from the effective fractal dimension as the thermodynamic limit of $\log(\widetilde{N}_q)=(D_q-\widetilde{D}_q)\log(N).$ Assuming ergodicity, this formula simplifies to $\log(\widetilde{N}_q)=(1-\widetilde{D}_q)\log(N)$ for $q>0$ and to  $\log(\widetilde{N}_0)=(-1+\widetilde{\alpha}_0)\log(N)$ for the logarithm of typical wavefunctions, which are the quantities shown in panel (a), (b) and (c) of Fig.\ \ref{fig:loglo_x1}, respectively. The solid lines are Padé approximants for each size data set. We have repeated the fitting procedure used for $D_1$ in Fig.\ \ref{fig:D1} for $\widetilde{D}_2$ and $\widetilde{\alpha}_0.$ Doing so, we have obtained extrapolated ergodic values $D_2=\alpha_0=1$ for all the disorders in which corrections Eq.~\ref{eq:corr} with $r=s=1$ produced good quality fittings. We have also extracted  $\log(N_q),$ as the coefficient $a_{01}$ in Eq.~\ref{eq:corr}, which appears in each of its corresponding panels of Fig.\ \ref{fig:loglo_x1} as $\star.$ We have included the ergodic volume $N_e$ (dashed line) extracted from Refs.\ \cite{Biroli2017,Tikhonov_Mirlin_2021}), see Appendix\ \ref{sec:app2}, and the GOE value (semi-dashed line).

At small disorder, all the finite-size data in Fig.\ \ref{fig:loglo_x1} converge to its corresponding thermodynamic limit extrapolation $\log(N_q)$ (denoted with $\star$). The converged values at small disorder fit well with the ones predicted by GOE, while at intermediate values of disorder $W\approx 10$ show a behaviour given by
{
\begin{align}
\log(N_q)= \exp(A_q+B_qW)    \label{eq:asymtotic}
\end{align}
with $B_1\approx 0.235\pm0.008, B_2=0.254\pm 0.002, B_0=0.222\pm0.005$ and $A_1\approx -1.2,\ A_2\approx -0.6, \ A_0\approx -0.81.$ We notice that the $B_q$ seems to be close for all the cases considered, but not the same.  Leaving apart for a moment the small difference in the $B_q$ factors, our results are compatible with a standard critical phenomenon---also with the supersymmetric formalism\ \cite{Tikhonov_Mirlin_2021}---where the logarithm of all those correlation lengths diverge with an exponent $\nu.$ Indeed, all the $N_q$ seem to be related by $\log(N_q)\sim c_{qq^\prime}\log(N_{q^\prime}),$ where the proportionality constant is related to the $A_q$ in Eq.\ref{eq:asymtotic} as $c_{q{q^\prime}}=e^{A_q-A_{q^\prime}}.$ Returning to the small differences between the $B_q$ factors, they imply that the $\log(N_q)$ factors are not proportional between them, which may be because we are not so close to the critical regime.}

In this intermediate regime, the slope in double logarithmic scale of the ergodic volume computed via population dynamics in Ref.\ \cite{Biroli2017} (dashed lines in all panels of Fig.\ \ref{fig:loglo_x1}) is close to the $B\approx 0.24,$ so that $\log(N_e)\sim  \log(N_q)$ in this regime. Even if we are obtaining results fully compatible with ergodicity $D_q=1$ for these intermediate disorders, it is not correct to refer to any of the $N_q$ as ergodic volumes, as they only set the system size needed to obtain small leading corrections for the corresponding $\widetilde{D}_q$ moment. Indeed, Eq.~\ref{eq:asymtotic} implies large differences between correlated volumes with different $q$ values. {  Notice that our findings are compatible with a standard critical phenomenon, also with the supersymmetric formalism\ \cite{Tikhonov_Mirlin_2021}, where the logarithm of all those correlation lengths diverge with any exponent $\nu,$ as all of the $N_q$ seems to be related by $\log(N_q)\sim c_{qq^\prime}\log(N_{q^\prime}),$ where we can relate the proportionality constant to the numbers $A_q$ in Eq.\ref{eq:asymtotic} as $c_{q_q^\prime}=e^{A_q-A_{q^\prime}}.$}

\begin{figure}[t!]
\includegraphics[width=1.\columnwidth]{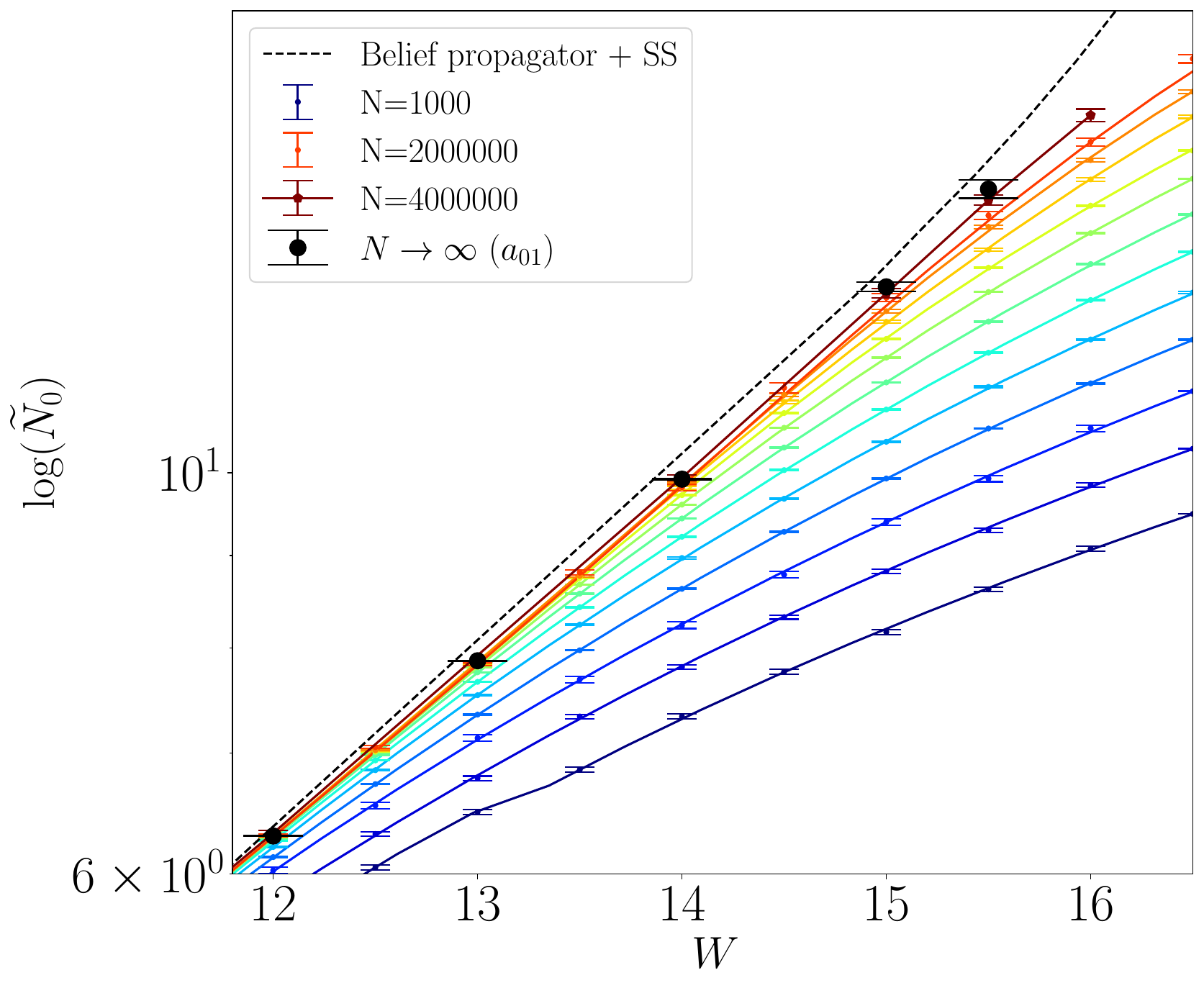}
\caption{\label{fig:N0} Finite-size corrections $\log(\widetilde{N}_0)$ assuming ergodicity as a function of disorder for several system sizes from $N=10^3$ to $N=4\times 10^6$. {  Notice the log scale in y-axe.} Solid lines are Padé approximants for each size but the largest for which is just a guide to the eye. The circles represent the extrapolated value of $\log(N_0)=a^{(0)}_{01}$ to the thermodynamic limit via a fitting $\widetilde{\alpha}_0 = \sum_{j=0,k=0}^{1} a^{(0)}_{jk}/(N^j[\log(N)]^k)$ where the ergodic value $a^{(0)}_{00}=\alpha_0=1$ is fixed. Several of the smaller sizes have been removed for each disorder so to obtain good quality fitting $p>1,$ for instance $5$ are used for $W=15.5,\ 16.$ The dashed line is the typical imaginary part of the self-energy obtained\ \cite{biroli2018delocalization, Tikhonov_Mirlin_2021}.
}
\end{figure}

Notice that $\widetilde{N}_0$ in Fig.\ \ref{fig:loglo_x1}(c) seems to have converged for larger disorders than the ones where we have reported its thermodynamic value. This encourages us to seek alternative fitting procedures to obtain that correlated volume closer to the transition. We have tried adding more corrections $r,s>1$ in Eq.~\ref{eq:corr}, but, although we could obtain good quality fittings doing so, it produces non-physical results in the parameter estimation, see Appendix\ \ref{supp:fittings}. Instead,  we have extracted the correlated volume $N_0$ via a fit of the data without taking into account the smallest sizes while fixing $r=s=1$ in Eq.~\ref{eq:corr}. The criteria to choose how many data points are included is based on the minimization of $|1-\chi_r^2|.$ Having a close to one $\chi^2_r$ we are sure that no over-fitting occurs. Additionally, we have fixed the ergodic value $\alpha_0=1$ in Eq.~\ref{eq:corr} to reduce the number of free parameters. See Appendix\ \ref{supp:fittings} for more information on these fittings and additional strategies to fit the data.

A zoom of $\log(\widetilde{N}_0)$ appears in Fig.\ \ref{fig:N0} for the finite system together with its extrapolated value $\log(N_0)$ (black scatters) up to disorder $W=15.5.$ Notice that the fit for this disorder only includes the four largest sizes, see Fig.\ \ref{Fig:alpha_f2} of Appendix, which produces a large error bar for the corresponding extrapolation. The analytical estimation $N_e$ (dashed line) is larger than our infinite size extrapolated $N_0$ for $W>13.$ The differences between both estimations become more pronounced at large disorder $W\approx 15,$ disorder up to which the tendency of $N_e$ begins to exhibit a tendency to diverge.  We cannot conclude whether the correlated volume  $\log(N_0)$ extracted from our numeric at $W=15.5$ is still described by Eq.~\ref{eq:asymtotic} or if it begins to deviate from that law due to the uncertainty in its extrapolation. In any case, no sign of a divergence at criticality can be inferred from our correlated volume. Thus, exact diagonalization results up to $N=4\times 10^6$ do not provide evidence of a divergence on the correlated volumes, specifically none with exponent $\nu=1/2,$ in contrast to Refs.\ \cite{Tikhonov_Mirlin_2021,tikhonov2019,mirlin1994distribution,biroli2018delocalization,Biroli2022}. {  We notice that the small differences in the slopes on log-log plot of correlated volumes versus disorder are here evident. Indeed, correlated volume $N_0$ shows a slope in double log scale that is definitely different than the analogous ones for the so-called ergodic volume $N_e.$}

\begin{figure}[t!]
\includegraphics[width=1.\columnwidth]{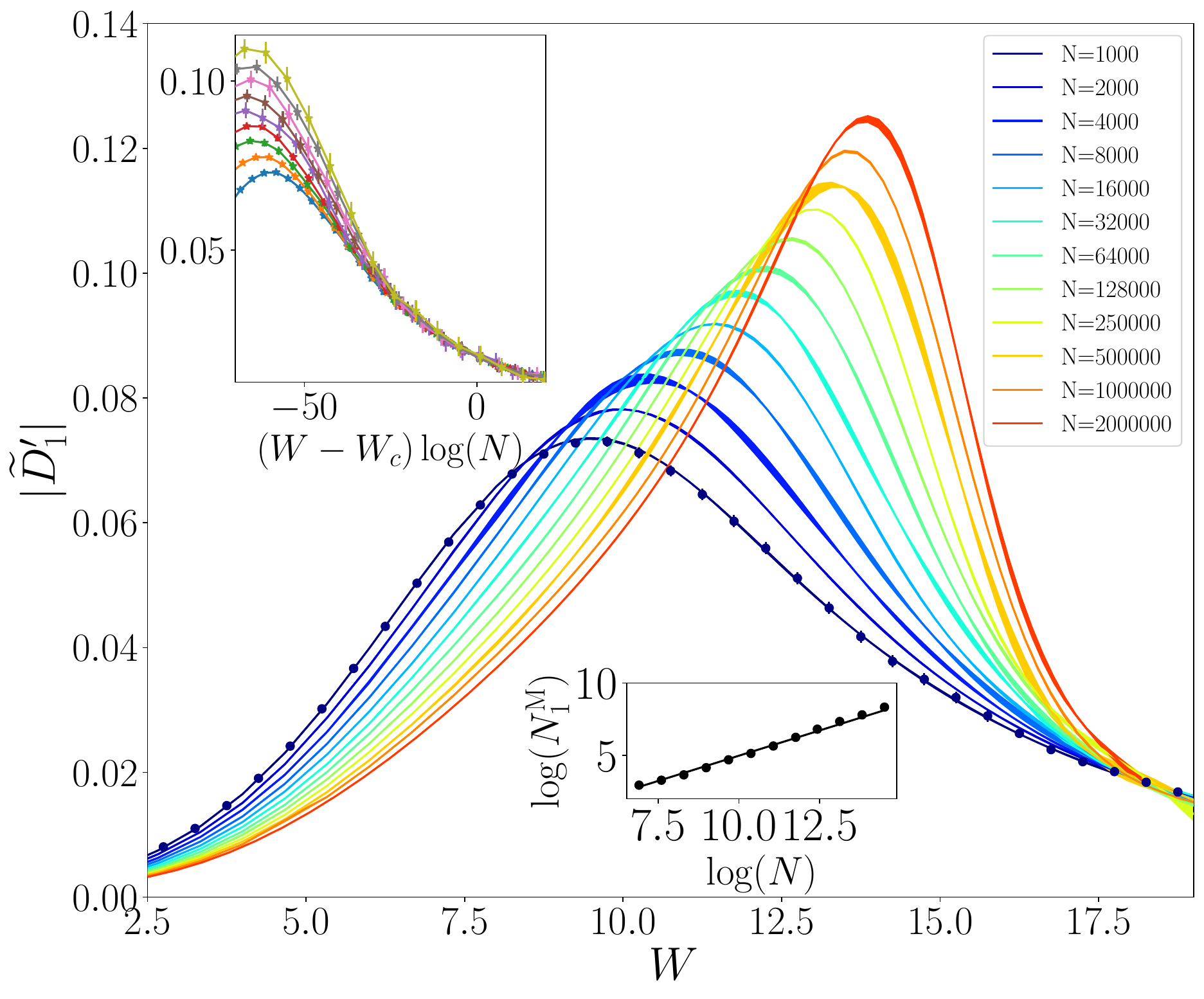}
\caption{\label{fig:D1p_pade} Absolute value of the derivative $\frac{d\widetilde{D}_1}{dW}$  as a function of disorder $W.$ Solid lines are the derivative of the Padé approximant for $\widetilde{D}_1$. The numerical derivative of $\widetilde{D}_1$ is included for the smallest size to check consistency with its corresponding Padé derivative. Top inset: numerical derivative of $\widetilde{D}_1$ near the Anderson transition $W_c\approx 18.17$ for sizes up to $N=128000$. Bottom inset: estimation of the logarithm of the first correlated volume $N_1^M=A_1+BW$ with $A_1=-1.2$ and $B=0.24$ (see Eq.~\ref{eq:asymtotic}) at the disorder where a maximum occurs at $\widetilde{D}_1$ for a given size $N$ (see main panel) as a function of the logarithm of system size. The solid line fits to $a \log(N) +b$ yielding $a\approx 0.7,\ b\approx -2$.  }
\end{figure}

\section{Finite-size effects near the Anderson transition}\label{Sec:nearAT}

As we have seen, our finite-size numerical results up to $N=4\times 10^6$ do not contain indications of divergences in $N_q,$ neither of a critical exponent $\nu=1/2$. One of us has reported additional evidence of a critical scaling in the derivative of fractal dimensions, not in the correlated volumes\ \cite{pino2020scaling}. In that work, the derivative of the first fractal dimension was found to display a crossing point for the curves corresponding to different sizes, so that a non-analyticity with a continuous $D_1$ and exponent $\nu=1$ seemed to develop at the Anderson transition. Here, we check that our data up to $N=250000$ are still compatible with such a scaling (error bars are too big for larger system sizes) and that this critical scaling only appears when the crossover to $N\ll N_1$ has taken place.

In Fig.\ \ref{fig:D1p_pade} we have plotted the derivative of Padé approximants of the first fractal dimension for lengths up to $N=2\times 10^6$. Maxima occur for all the sizes at intermediate values of disorder from $W=10$ to $W=15,$ around the same disorder where finite size-effects begin to be important in $\log( \widetilde{N}_1)$ in  Fig.\ \ref{fig:loglo_x1}(a). We can extract the correlated volume at each of the maxima $N_1^M(N)$ using Eq.\ \ref{eq:asymtotic} with the disorder $W_M(N)$ at which the maxima occur for a given size $N.$  Doing so we have been able to check that the maxima occurs when  $\log(N_1^M)\sim 0.7 \log(N),$ as shown at the bottom inset of Fig.\ \ref{fig:D1p_pade}.  Taking the number of nodes in the RRG as the size $L\sim \log(N),$ this implies that the crossover is controlled by the diameter of the graph $L\sim \log(N),$ not by its volume\ \cite{cardy1996scaling}.   Such a crossover occurs for the correlated volumes because they grow very fast Eq.~\ref{eq:corr}, even if there is not evidence of a divergence as in a standard second-order phase transition.

Besides the crossover we have already commented on, there is an additional critical behaviour in a RRG with exponent $\nu=1.$ This can be seen in the upper inset of Fig\ \ref{fig:D1p_pade}, where all the curves for the first fractal dimension collapse when plotted as a function of the scaling variable $\log(N)(W-W_c)^{1/\nu}$ with $\nu=1.$ {  We have not attempted to fit our data to a scaling form as in Ref.\ \cite{pino2020scaling} because the data $D_1^{\prime}$for the larger sizes do not contain enough number of disorder realizations as to give accurate results.} Notice that the region of this critical scaling---the region where the curves for different sizes collapse into a single curve---never occurs for disorder around the maxima. Thus, the additional critical scaling with $\nu=1$ only occurs once the crossover has taken place.

\section{Summary and conclusions}

{  We have analyzed the Anderson problem in a RRG with new quantities---the correlated volumes---, taking advantage of new polynomial filters implemented in SLEPC. Using them, we have obtained results compatible with ergodicity $D_q=1$ and correlated volumes growing fast $\log(\log(N_q))\sim W/4$ in an intermediate metallic regime. The zero correlated volume is similar to the ergodic volume $N_e$ found from the typical value of the local density of states in Refs.\ \cite{Tikhonov_Mirlin_2021,biroli2018delocalization,Biroli2022}.} We have shown further evidence of a critical behavior with $\nu=1$ in $\widetilde{D}^\prime_1$ once the crossover $\log(N)\ll \log(N_1)$ has taken place. As we have not obtained evidence of divergence on correlated volumes,  we cannot validate the picture of a critical exponent $\nu=1/2$ derived from the supersymmetric formalism\ \cite{Tikhonov_Mirlin_2021,mirlin1994distribution, Mirlin_Fyodorov_1991}. Taking into account that our system sizes are much larger than the ones used in previous works, previous evidence of such a divergence from exact diagonalization results should be revised.

{  Several scenarios are compatible with our data. The most standard one is that the self-consistent approximation for the probability distribution of the Greens function holds. Then, a divergence of correlated volumes can occur with $\nu=1/2$ for larger disorder values than the one for which we can reliably extract correlated volumes, $W\approx 15.$ If this is the case, the situation will be fully described by the supersymmetric formalism and, as a result, the full metal may become ergodic\ \cite{mirlin1994distribution,tikhonov2018,tikhonov2019,Tikhonov_Mirlin_2021}.} Notice that, in this case, the additional critical behaviour in $D_1^\prime$ near the Anderson transition, see Sec.\ \ref{Sec:nearAT}, may be a consequence of the Bethe- and Random-Regular- graphs equivalence for sizes much smaller than the first correlated volume, similarly as discussed in Ref.\ \cite{Tikhonov2016}. Then, we can talk about an ergodic diameter, not ergodic volume, as $\log(N_e)$ marks the scale of graph diameter from which finite size corrections to the thermodynamic result vanishes in all the $I_q.$ { Even in this scenario, there are still details that one needs to understand better, such as the differences between the slopes of correlated volumes in double logarithmic scale as a function of disorder strength for different $q$, see Eq.\ \ref{eq:asymtotic} and Figs.\ \ref{fig:loglo_x1},\ref{fig:N0}. As a consequence, notice the differences between the thermodynamic values of the correlated volume $N_0$ (bullets) and the ergodic volume computed from the self-consistent solution (dashed line) in Fig.\ \ref{fig:N0}. Closed loops in the self-consistent equations may help to clarify this \cite{baroni2023corrections}.}

Another scenario compatible with our numerical data is that at disorder $15<W<W_c$ there is a genuine non-ergodic regime with some of the fractal dimension becoming smaller than its ergodic value $D_q<1.$ Notice that none of our results for $\log(\widetilde{N}_0)$ in Fig.\ \ref{fig:loglo_x1} show a clear convergence with system size for disorder $W>15.5.$  This may be an indication that $\alpha_0>1$ has a non-ergodic value in the thermodynamic limit, which implies $\log(\widetilde{N}_0)\sim \log(N).$ This is an interesting case worth analyzing further. It would imply that the self-consistent approximation that leads to the analytical supersymmetric formula\ \cite{mirlin1994distribution,Tikhonov_Mirlin_2021}  does not capture the multifractal character of the wavefunctions. This situation may be similar to the one that occurred in spin-glasses on RRG twenty years ago\ \cite{mezard2001bethe}, see\ \cite{kravtsov2018non}.

Despite we have analyzed large graphs, an effort should be made to go even for larger ones. We are confident that our numerical method will still work for sizes $N=8\times 10^6$ or even larger. Besides disordered RRGs, we believe that the numerical methods together with the improved finite-size analysis employed here will be beneficial in a large class of problems involving disordered systems.

\begin{acknowledgments}
We thank Alberto Rodríguez for useful discussions. M.~P.\ and J.~E.~R.\ acknowledge support by Spanish MCIN/AEI/10.13039/501100011033 through Grants PID2020-114830GB-I00 and PID2022-139568NB-I00, respectively. M.~P.\ further acknowledges funding by European Commission FET-Open project AVaQus GA 899561 and Consolidación Investigadora CNS2022-136025. The computational experiments were carried out on the supercomputer Tirant III belonging to Universitat de Val\`encia and SCAYLE, Supercomputación Castilla y León.
\end{acknowledgments}

\bibliography{./MBLandNonequilibrium}

\begin{thebibliography}{53}%
\makeatletter
\providecommand \@ifxundefined [1]{%
 \@ifx{#1\undefined}
}%
\providecommand \@ifnum [1]{%
 \ifnum #1\expandafter \@firstoftwo
 \else \expandafter \@secondoftwo
 \fi
}%
\providecommand \@ifx [1]{%
 \ifx #1\expandafter \@firstoftwo
 \else \expandafter \@secondoftwo
 \fi
}%
\providecommand \natexlab [1]{#1}%
\providecommand \enquote  [1]{``#1''}%
\providecommand \bibnamefont  [1]{#1}%
\providecommand \bibfnamefont [1]{#1}%
\providecommand \citenamefont [1]{#1}%
\providecommand \href@noop [0]{\@secondoftwo}%
\providecommand \href [0]{\begingroup \@sanitize@url \@href}%
\providecommand \@href[1]{\@@startlink{#1}\@@href}%
\providecommand \@@href[1]{\endgroup#1\@@endlink}%
\providecommand \@sanitize@url [0]{\catcode `\\12\catcode `\$12\catcode
  `\&12\catcode `\#12\catcode `\^12\catcode `\_12\catcode `\%12\relax}%
\providecommand \@@startlink[1]{}%
\providecommand \@@endlink[0]{}%
\providecommand \url  [0]{\begingroup\@sanitize@url \@url }%
\providecommand \@url [1]{\endgroup\@href {#1}{\urlprefix }}%
\providecommand \urlprefix  [0]{URL }%
\providecommand \Eprint [0]{\href }%
\providecommand \doibase [0]{https://doi.org/}%
\providecommand \selectlanguage [0]{\@gobble}%
\providecommand \bibinfo  [0]{\@secondoftwo}%
\providecommand \bibfield  [0]{\@secondoftwo}%
\providecommand \translation [1]{[#1]}%
\providecommand \BibitemOpen [0]{}%
\providecommand \bibitemStop [0]{}%
\providecommand \bibitemNoStop [0]{.\EOS\space}%
\providecommand \EOS [0]{\spacefactor3000\relax}%
\providecommand \BibitemShut  [1]{\csname bibitem#1\endcsname}%
\let\auto@bib@innerbib\@empty
\bibitem [{\citenamefont {Aizenman}\ and\ \citenamefont
  {Warzel}(2011)}]{aizenman2011extended}%
  \BibitemOpen
  \bibfield  {author} {\bibinfo {author} {\bibfnamefont {M.}~\bibnamefont
  {Aizenman}}\ and\ \bibinfo {author} {\bibfnamefont {S.}~\bibnamefont
  {Warzel}},\ }\bibfield  {title} {\bibinfo {title} {Extended states in a
  lifshitz tail regime for random schr{\"o}dinger operators on trees},\
  }\href@noop {} {\bibfield  {journal} {\bibinfo  {journal} {Physical review
  letters}\ }\textbf {\bibinfo {volume} {106}},\ \bibinfo {pages} {136804}
  (\bibinfo {year} {2011})}\BibitemShut {NoStop}%
\bibitem [{\citenamefont {Warzel}\ and\ \citenamefont
  {Aizenman}(2013)}]{warzel2013resonant}%
  \BibitemOpen
  \bibfield  {author} {\bibinfo {author} {\bibfnamefont {S.}~\bibnamefont
  {Warzel}}\ and\ \bibinfo {author} {\bibfnamefont {M.}~\bibnamefont
  {Aizenman}},\ }\bibfield  {title} {\bibinfo {title} {Resonant delocalization
  for random schr{\"o}dinger operators on tree graphs},\ }\href@noop {}
  {\bibfield  {journal} {\bibinfo  {journal} {Journal of the European
  Mathematical Society}\ }\textbf {\bibinfo {volume} {15}},\ \bibinfo {pages}
  {1167} (\bibinfo {year} {2013})}\BibitemShut {NoStop}%
\bibitem [{\citenamefont {Kravtsov}\ \emph {et~al.}(2015)\citenamefont
  {Kravtsov}, \citenamefont {Khaymovich}, \citenamefont {Cuevas},\ and\
  \citenamefont {Amini}}]{Kravtsov2015}%
  \BibitemOpen
  \bibfield  {author} {\bibinfo {author} {\bibfnamefont {V.~E.}\ \bibnamefont
  {Kravtsov}}, \bibinfo {author} {\bibfnamefont {I.~M.}\ \bibnamefont
  {Khaymovich}}, \bibinfo {author} {\bibfnamefont {E.}~\bibnamefont {Cuevas}},\
  and\ \bibinfo {author} {\bibfnamefont {M.}~\bibnamefont {Amini}},\ }\bibfield
   {title} {\bibinfo {title} {A random matrix model with localization and
  ergodic transitions},\ }\href
  {http://stacks.iop.org/1367-2630/17/i=12/a=122002} {\bibfield  {journal}
  {\bibinfo  {journal} {New Journal of Physics}\ }\textbf {\bibinfo {volume}
  {17}},\ \bibinfo {pages} {122002} (\bibinfo {year} {2015})}\BibitemShut
  {NoStop}%
\bibitem [{\citenamefont {Khaymovich}\ and\ \citenamefont
  {Kravtsov}(2021)}]{khaymovich2021dynamical}%
  \BibitemOpen
  \bibfield  {author} {\bibinfo {author} {\bibfnamefont {I.}~\bibnamefont
  {Khaymovich}}\ and\ \bibinfo {author} {\bibfnamefont {V.}~\bibnamefont
  {Kravtsov}},\ }\bibfield  {title} {\bibinfo {title} {Dynamical phases in
  a``multifractal''rosenzweig-porter model},\ }\href@noop {} {\bibfield
  {journal} {\bibinfo  {journal} {SciPost Physics}\ }\textbf {\bibinfo {volume}
  {11}},\ \bibinfo {pages} {045} (\bibinfo {year} {2021})}\BibitemShut
  {NoStop}%
\bibitem [{\citenamefont {Pino}\ \emph {et~al.}(2019)\citenamefont {Pino},
  \citenamefont {Tabanera},\ and\ \citenamefont {Serna}}]{pino2019ergodic}%
  \BibitemOpen
  \bibfield  {author} {\bibinfo {author} {\bibfnamefont {M.}~\bibnamefont
  {Pino}}, \bibinfo {author} {\bibfnamefont {J.}~\bibnamefont {Tabanera}},\
  and\ \bibinfo {author} {\bibfnamefont {P.}~\bibnamefont {Serna}},\ }\bibfield
   {title} {\bibinfo {title} {From ergodic to non-ergodic chaos in
  rosenzweig--porter model},\ }\href@noop {} {\bibfield  {journal} {\bibinfo
  {journal} {Journal of Physics A: Mathematical and Theoretical}\ }\textbf
  {\bibinfo {volume} {52}},\ \bibinfo {pages} {475101} (\bibinfo {year}
  {2019})}\BibitemShut {NoStop}%
\bibitem [{\citenamefont {Altshuler}\ \emph {et~al.}(1997)\citenamefont
  {Altshuler}, \citenamefont {Gefen}, \citenamefont {Kamenev},\ and\
  \citenamefont {Levitov}}]{Al1997}%
  \BibitemOpen
  \bibfield  {author} {\bibinfo {author} {\bibfnamefont {B.~L.}\ \bibnamefont
  {Altshuler}}, \bibinfo {author} {\bibfnamefont {Y.}~\bibnamefont {Gefen}},
  \bibinfo {author} {\bibfnamefont {A.}~\bibnamefont {Kamenev}},\ and\ \bibinfo
  {author} {\bibfnamefont {L.~S.}\ \bibnamefont {Levitov}},\ }\bibfield
  {title} {\bibinfo {title} {Quasiparticle lifetime in a finite system: A
  nonperturbative approach},\ }\href
  {https://doi.org/10.1103/PhysRevLett.78.2803} {\bibfield  {journal} {\bibinfo
   {journal} {Phys. Rev. Lett.}\ }\textbf {\bibinfo {volume} {78}},\ \bibinfo
  {pages} {2803} (\bibinfo {year} {1997})}\BibitemShut {NoStop}%
\bibitem [{\citenamefont {Pino}\ \emph {et~al.}(2016)\citenamefont {Pino},
  \citenamefont {Ioffe},\ and\ \citenamefont {Altshuler}}]{Pino15}%
  \BibitemOpen
  \bibfield  {author} {\bibinfo {author} {\bibfnamefont {M.}~\bibnamefont
  {Pino}}, \bibinfo {author} {\bibfnamefont {L.~B.}\ \bibnamefont {Ioffe}},\
  and\ \bibinfo {author} {\bibfnamefont {B.~L.}\ \bibnamefont {Altshuler}},\
  }\bibfield  {title} {\bibinfo {title} {Nonergodic metallic and insulating
  phases of josephson junction chains},\ }\href@noop {} {\bibfield  {journal}
  {\bibinfo  {journal} {Proceedings of the National Academy of Sciences}\
  }\textbf {\bibinfo {volume} {113}},\ \bibinfo {pages} {536} (\bibinfo {year}
  {2016})}\BibitemShut {NoStop}%
\bibitem [{\citenamefont {Venturelli}\ \emph {et~al.}(2023)\citenamefont
  {Venturelli}, \citenamefont {Cugliandolo}, \citenamefont {Schehr},\ and\
  \citenamefont {Tarzia}}]{venturelli2022replica}%
  \BibitemOpen
  \bibfield  {author} {\bibinfo {author} {\bibfnamefont {D.}~\bibnamefont
  {Venturelli}}, \bibinfo {author} {\bibfnamefont {L.~F.}\ \bibnamefont
  {Cugliandolo}}, \bibinfo {author} {\bibfnamefont {G.}~\bibnamefont
  {Schehr}},\ and\ \bibinfo {author} {\bibfnamefont {M.}~\bibnamefont
  {Tarzia}},\ }\bibfield  {title} {\bibinfo {title} {Replica approach to the
  generalized rosenzweig-porter model},\ }\href@noop {} {\bibfield  {journal}
  {\bibinfo  {journal} {SciPost Physics}\ }\textbf {\bibinfo {volume} {14}},\
  \bibinfo {pages} {110} (\bibinfo {year} {2023})}\BibitemShut {NoStop}%
\bibitem [{\citenamefont {Tikhonov}\ \emph {et~al.}(2016)\citenamefont
  {Tikhonov}, \citenamefont {Mirlin},\ and\ \citenamefont
  {Skvortsov}}]{Tikhonov2016b}%
  \BibitemOpen
  \bibfield  {author} {\bibinfo {author} {\bibfnamefont {K.~S.}\ \bibnamefont
  {Tikhonov}}, \bibinfo {author} {\bibfnamefont {A.~D.}\ \bibnamefont
  {Mirlin}},\ and\ \bibinfo {author} {\bibfnamefont {M.~A.}\ \bibnamefont
  {Skvortsov}},\ }\bibfield  {title} {\bibinfo {title} {Anderson localization
  on random regular graphs},\ }\href@noop {} {\bibfield  {journal} {\bibinfo
  {journal} {Physical Review B}\ }\textbf {\bibinfo {volume} {94}},\ \bibinfo
  {pages} {220203(R)} (\bibinfo {year} {2016})}\BibitemShut {NoStop}%
\bibitem [{\citenamefont {Tikhonov}\ and\ \citenamefont
  {Mirlin}(2019)}]{tikhonov2019}%
  \BibitemOpen
  \bibfield  {author} {\bibinfo {author} {\bibfnamefont {K.~S.}\ \bibnamefont
  {Tikhonov}}\ and\ \bibinfo {author} {\bibfnamefont {A.~D.}\ \bibnamefont
  {Mirlin}},\ }\bibfield  {title} {\bibinfo {title} {Critical behavior at the
  localization transition on random regular graphs},\ }\href
  {https://doi.org/10.1103/PhysRevB.99.214202} {\bibfield  {journal} {\bibinfo
  {journal} {Phys. Rev. B}\ }\textbf {\bibinfo {volume} {99}},\ \bibinfo
  {pages} {214202} (\bibinfo {year} {2019})}\BibitemShut {NoStop}%
\bibitem [{\citenamefont {Dorogovtsev}\ \emph {et~al.}(2008)\citenamefont
  {Dorogovtsev}, \citenamefont {Goltsev},\ and\ \citenamefont
  {Mendes}}]{dorogovtsev2008critical}%
  \BibitemOpen
  \bibfield  {author} {\bibinfo {author} {\bibfnamefont {S.~N.}\ \bibnamefont
  {Dorogovtsev}}, \bibinfo {author} {\bibfnamefont {A.~V.}\ \bibnamefont
  {Goltsev}},\ and\ \bibinfo {author} {\bibfnamefont {J.~F.}\ \bibnamefont
  {Mendes}},\ }\bibfield  {title} {\bibinfo {title} {Critical phenomena in
  complex networks},\ }\href@noop {} {\bibfield  {journal} {\bibinfo  {journal}
  {Reviews of Modern Physics}\ }\textbf {\bibinfo {volume} {80}},\ \bibinfo
  {pages} {1275} (\bibinfo {year} {2008})}\BibitemShut {NoStop}%
\bibitem [{\citenamefont {M{\'e}zard}\ and\ \citenamefont
  {Parisi}(2001)}]{mezard2001bethe}%
  \BibitemOpen
  \bibfield  {author} {\bibinfo {author} {\bibfnamefont {M.}~\bibnamefont
  {M{\'e}zard}}\ and\ \bibinfo {author} {\bibfnamefont {G.}~\bibnamefont
  {Parisi}},\ }\bibfield  {title} {\bibinfo {title} {The bethe lattice spin
  glass revisited},\ }\href@noop {} {\bibfield  {journal} {\bibinfo  {journal}
  {The European Physical Journal B-Condensed Matter and Complex Systems}\
  }\textbf {\bibinfo {volume} {20}},\ \bibinfo {pages} {217} (\bibinfo {year}
  {2001})}\BibitemShut {NoStop}%
\bibitem [{\citenamefont {Basko}\ \emph {et~al.}(2006)\citenamefont {Basko},
  \citenamefont {Aleiner},\ and\ \citenamefont {Altshuler}}]{Basko2006}%
  \BibitemOpen
  \bibfield  {author} {\bibinfo {author} {\bibfnamefont {D.}~\bibnamefont
  {Basko}}, \bibinfo {author} {\bibfnamefont {I.}~\bibnamefont {Aleiner}},\
  and\ \bibinfo {author} {\bibfnamefont {B.}~\bibnamefont {Altshuler}},\
  }\bibfield  {title} {\bibinfo {title} {Metal-insulator transition in a weakly
  interacting many-electron system with localized single-particle states},\
  }\href {http://dx.doi.org/10.1016/j.aop.2005.11.014} {\bibfield  {journal}
  {\bibinfo  {journal} {Annals of Physics}\ }\textbf {\bibinfo {volume}
  {321}},\ \bibinfo {pages} {1126} (\bibinfo {year} {2006})}\BibitemShut
  {NoStop}%
\bibitem [{\citenamefont {Altshuler}\ \emph {et~al.}(2010)\citenamefont
  {Altshuler}, \citenamefont {Krovi},\ and\ \citenamefont
  {Roland}}]{altshuler2010anderson}%
  \BibitemOpen
  \bibfield  {author} {\bibinfo {author} {\bibfnamefont {B.}~\bibnamefont
  {Altshuler}}, \bibinfo {author} {\bibfnamefont {H.}~\bibnamefont {Krovi}},\
  and\ \bibinfo {author} {\bibfnamefont {J.}~\bibnamefont {Roland}},\
  }\bibfield  {title} {\bibinfo {title} {Anderson localization makes adiabatic
  quantum optimization fail},\ }\href@noop {} {\bibfield  {journal} {\bibinfo
  {journal} {Proceedings of the National Academy of Sciences}\ }\textbf
  {\bibinfo {volume} {107}},\ \bibinfo {pages} {12446} (\bibinfo {year}
  {2010})}\BibitemShut {NoStop}%
\bibitem [{\citenamefont {Laumann}\ \emph {et~al.}(2015)\citenamefont
  {Laumann}, \citenamefont {Moessner}, \citenamefont {Scardicchio},\ and\
  \citenamefont {Sondhi}}]{laumann2015quantum}%
  \BibitemOpen
  \bibfield  {author} {\bibinfo {author} {\bibfnamefont {C.~R.}\ \bibnamefont
  {Laumann}}, \bibinfo {author} {\bibfnamefont {R.}~\bibnamefont {Moessner}},
  \bibinfo {author} {\bibfnamefont {A.}~\bibnamefont {Scardicchio}},\ and\
  \bibinfo {author} {\bibfnamefont {S.}~\bibnamefont {Sondhi}},\ }\bibfield
  {title} {\bibinfo {title} {Quantum annealing: The fastest route to quantum
  computation?},\ }\href@noop {} {\bibfield  {journal} {\bibinfo  {journal}
  {The European Physical Journal Special Topics}\ }\textbf {\bibinfo {volume}
  {224}},\ \bibinfo {pages} {75} (\bibinfo {year} {2015})}\BibitemShut
  {NoStop}%
\bibitem [{\citenamefont {Smelyanskiy}\ \emph {et~al.}(2020)\citenamefont
  {Smelyanskiy}, \citenamefont {Kechedzhi}, \citenamefont {Boixo},
  \citenamefont {Isakov}, \citenamefont {Neven},\ and\ \citenamefont
  {Altshuler}}]{smelyanskiy2020nonergodic}%
  \BibitemOpen
  \bibfield  {author} {\bibinfo {author} {\bibfnamefont {V.~N.}\ \bibnamefont
  {Smelyanskiy}}, \bibinfo {author} {\bibfnamefont {K.}~\bibnamefont
  {Kechedzhi}}, \bibinfo {author} {\bibfnamefont {S.}~\bibnamefont {Boixo}},
  \bibinfo {author} {\bibfnamefont {S.~V.}\ \bibnamefont {Isakov}}, \bibinfo
  {author} {\bibfnamefont {H.}~\bibnamefont {Neven}},\ and\ \bibinfo {author}
  {\bibfnamefont {B.}~\bibnamefont {Altshuler}},\ }\bibfield  {title} {\bibinfo
  {title} {Nonergodic delocalized states for efficient population transfer
  within a narrow band of the energy landscape},\ }\href@noop {} {\bibfield
  {journal} {\bibinfo  {journal} {Physical Review X}\ }\textbf {\bibinfo
  {volume} {10}},\ \bibinfo {pages} {011017} (\bibinfo {year}
  {2020})}\BibitemShut {NoStop}%
\bibitem [{\citenamefont {Faoro}\ \emph {et~al.}(2019)\citenamefont {Faoro},
  \citenamefont {Feigel'man},\ and\ \citenamefont {Ioffe}}]{faoro2018non}%
  \BibitemOpen
  \bibfield  {author} {\bibinfo {author} {\bibfnamefont {L.}~\bibnamefont
  {Faoro}}, \bibinfo {author} {\bibfnamefont {M.~V.}\ \bibnamefont
  {Feigel'man}},\ and\ \bibinfo {author} {\bibfnamefont {L.}~\bibnamefont
  {Ioffe}},\ }\bibfield  {title} {\bibinfo {title} {Non-ergodic extended phase
  of the quantum random energy model},\ }\href@noop {} {\bibfield  {journal}
  {\bibinfo  {journal} {Annals of Physics}\ }\textbf {\bibinfo {volume}
  {409}},\ \bibinfo {pages} {167916} (\bibinfo {year} {2019})}\BibitemShut
  {NoStop}%
\bibitem [{\citenamefont {Pino}\ and\ \citenamefont
  {Garcia-Ripoll}(2020)}]{Pino_Ripoll_2020}%
  \BibitemOpen
  \bibfield  {author} {\bibinfo {author} {\bibfnamefont {M.}~\bibnamefont
  {Pino}}\ and\ \bibinfo {author} {\bibfnamefont {J.~J.}\ \bibnamefont
  {Garcia-Ripoll}},\ }\bibfield  {title} {\bibinfo {title} {Mediator-assisted
  cooling in quantum annealing},\ }\href
  {https://doi.org/10.1103/PhysRevA.101.032324} {\bibfield  {journal} {\bibinfo
   {journal} {Physical Review A}\ }\textbf {\bibinfo {volume} {101}},\ \bibinfo
  {pages} {032324} (\bibinfo {year} {2020})}\BibitemShut {NoStop}%
\bibitem [{\citenamefont {Pino}\ \emph {et~al.}(2013)\citenamefont {Pino},
  \citenamefont {Prior},\ and\ \citenamefont {Clark}}]{pino2013capturing}%
  \BibitemOpen
  \bibfield  {author} {\bibinfo {author} {\bibfnamefont {M.}~\bibnamefont
  {Pino}}, \bibinfo {author} {\bibfnamefont {J.}~\bibnamefont {Prior}},\ and\
  \bibinfo {author} {\bibfnamefont {S.}~\bibnamefont {Clark}},\ }\bibfield
  {title} {\bibinfo {title} {Capturing the re-entrant behavior of
  one-dimensional bose--hubbard model},\ }\href@noop {} {\bibfield  {journal}
  {\bibinfo  {journal} {physica status solidi (b)}\ }\textbf {\bibinfo {volume}
  {250}},\ \bibinfo {pages} {51} (\bibinfo {year} {2013})}\BibitemShut
  {NoStop}%
\bibitem [{\citenamefont {De~Luca}\ \emph {et~al.}(2014)\citenamefont
  {De~Luca}, \citenamefont {Altshuler}, \citenamefont {Kravtsov},\ and\
  \citenamefont {Scardicchio}}]{Deluca2014}%
  \BibitemOpen
  \bibfield  {author} {\bibinfo {author} {\bibfnamefont {A.}~\bibnamefont
  {De~Luca}}, \bibinfo {author} {\bibfnamefont {B.~L.}\ \bibnamefont
  {Altshuler}}, \bibinfo {author} {\bibfnamefont {V.~E.}\ \bibnamefont
  {Kravtsov}},\ and\ \bibinfo {author} {\bibfnamefont {A.}~\bibnamefont
  {Scardicchio}},\ }\bibfield  {title} {\bibinfo {title} {Anderson localization
  on the bethe lattice: Nonergodicity of extended states},\ }\href
  {https://doi.org/10.1103/PhysRevLett.113.046806} {\bibfield  {journal}
  {\bibinfo  {journal} {Phys. Rev. Lett.}\ }\textbf {\bibinfo {volume} {113}},\
  \bibinfo {pages} {046806} (\bibinfo {year} {2014})}\BibitemShut {NoStop}%
\bibitem [{\citenamefont {Cuevas}\ \emph {et~al.}(2001)\citenamefont {Cuevas},
  \citenamefont {Gasparian},\ and\ \citenamefont {Ortu\~no}}]{Cuevas2001}%
  \BibitemOpen
  \bibfield  {author} {\bibinfo {author} {\bibfnamefont {E.}~\bibnamefont
  {Cuevas}}, \bibinfo {author} {\bibfnamefont {V.}~\bibnamefont {Gasparian}},\
  and\ \bibinfo {author} {\bibfnamefont {M.}~\bibnamefont {Ortu\~no}},\
  }\bibfield  {title} {\bibinfo {title} {Anomalously large critical regions in
  power-law random matrix ensembles},\ }\href
  {https://doi.org/10.1103/PhysRevLett.87.056601} {\bibfield  {journal}
  {\bibinfo  {journal} {Phys. Rev. Lett.}\ }\textbf {\bibinfo {volume} {87}},\
  \bibinfo {pages} {056601} (\bibinfo {year} {2001})}\BibitemShut {NoStop}%
\bibitem [{\citenamefont {Kravtsov}\ \emph {et~al.}(2018)\citenamefont
  {Kravtsov}, \citenamefont {Altshuler},\ and\ \citenamefont
  {Ioffe}}]{kravtsov2018non}%
  \BibitemOpen
  \bibfield  {author} {\bibinfo {author} {\bibfnamefont {V.}~\bibnamefont
  {Kravtsov}}, \bibinfo {author} {\bibfnamefont {B.}~\bibnamefont
  {Altshuler}},\ and\ \bibinfo {author} {\bibfnamefont {L.}~\bibnamefont
  {Ioffe}},\ }\bibfield  {title} {\bibinfo {title} {Non-ergodic delocalized
  phase in anderson model on bethe lattice and regular graph},\ }\href@noop {}
  {\bibfield  {journal} {\bibinfo  {journal} {Annals of Physics}\ }\textbf
  {\bibinfo {volume} {389}},\ \bibinfo {pages} {148} (\bibinfo {year}
  {2018})}\BibitemShut {NoStop}%
\bibitem [{\citenamefont {Biroli}\ and\ \citenamefont
  {Tarzia}(2020)}]{biroli2020anomalous}%
  \BibitemOpen
  \bibfield  {author} {\bibinfo {author} {\bibfnamefont {G.}~\bibnamefont
  {Biroli}}\ and\ \bibinfo {author} {\bibfnamefont {M.}~\bibnamefont
  {Tarzia}},\ }\bibfield  {title} {\bibinfo {title} {Anomalous dynamics on the
  ergodic side of the many-body localization transition and the glassy phase of
  directed polymers in random media},\ }\href
  {https://doi.org/10.1103/PhysRevB.102.064211} {\bibfield  {journal} {\bibinfo
   {journal} {Phys. Rev. B}\ }\textbf {\bibinfo {volume} {102}},\ \bibinfo
  {pages} {064211} (\bibinfo {year} {2020})}\BibitemShut {NoStop}%
\bibitem [{\citenamefont {Colmenarez}\ \emph {et~al.}(2022)\citenamefont
  {Colmenarez}, \citenamefont {Luitz}, \citenamefont {Khaymovich},\ and\
  \citenamefont {De~Tomasi}}]{colmenarez2022}%
  \BibitemOpen
  \bibfield  {author} {\bibinfo {author} {\bibfnamefont {L.}~\bibnamefont
  {Colmenarez}}, \bibinfo {author} {\bibfnamefont {D.~J.}\ \bibnamefont
  {Luitz}}, \bibinfo {author} {\bibfnamefont {I.~M.}\ \bibnamefont
  {Khaymovich}},\ and\ \bibinfo {author} {\bibfnamefont {G.}~\bibnamefont
  {De~Tomasi}},\ }\bibfield  {title} {\bibinfo {title} {Subdiffusive thouless
  time scaling in the anderson model on random regular graphs},\ }\href
  {https://doi.org/10.1103/PhysRevB.105.174207} {\bibfield  {journal} {\bibinfo
   {journal} {Phys. Rev. B}\ }\textbf {\bibinfo {volume} {105}},\ \bibinfo
  {pages} {174207} (\bibinfo {year} {2022})}\BibitemShut {NoStop}%
\bibitem [{\citenamefont {Tikhonov}\ and\ \citenamefont
  {Mirlin}(2016)}]{Tikhonov2016}%
  \BibitemOpen
  \bibfield  {author} {\bibinfo {author} {\bibfnamefont {K.~S.}\ \bibnamefont
  {Tikhonov}}\ and\ \bibinfo {author} {\bibfnamefont {A.~D.}\ \bibnamefont
  {Mirlin}},\ }\bibfield  {title} {\bibinfo {title} {Fractality of wave
  functions on a cayley tree: Difference between tree and locally treelike
  graph without boundary},\ }\href {https://doi.org/10.1103/PhysRevB.94.184203}
  {\bibfield  {journal} {\bibinfo  {journal} {Phys. Rev. B}\ }\textbf {\bibinfo
  {volume} {94}},\ \bibinfo {pages} {184203} (\bibinfo {year}
  {2016})}\BibitemShut {NoStop}%
\bibitem [{\citenamefont {Garcia-Mata}\ \emph {et~al.}(2017)\citenamefont
  {Garcia-Mata}, \citenamefont {Giraud}, \citenamefont {Georgeot},
  \citenamefont {Martin}, \citenamefont {Dubertrand},\ and\ \citenamefont
  {Lemari{\'e}}}]{garcia2017scaling}%
  \BibitemOpen
  \bibfield  {author} {\bibinfo {author} {\bibfnamefont {I.}~\bibnamefont
  {Garcia-Mata}}, \bibinfo {author} {\bibfnamefont {O.}~\bibnamefont {Giraud}},
  \bibinfo {author} {\bibfnamefont {B.}~\bibnamefont {Georgeot}}, \bibinfo
  {author} {\bibfnamefont {J.}~\bibnamefont {Martin}}, \bibinfo {author}
  {\bibfnamefont {R.}~\bibnamefont {Dubertrand}},\ and\ \bibinfo {author}
  {\bibfnamefont {G.}~\bibnamefont {Lemari{\'e}}},\ }\bibfield  {title}
  {\bibinfo {title} {Scaling theory of the anderson transition in random
  graphs: ergodicity and universality},\ }\href@noop {} {\bibfield  {journal}
  {\bibinfo  {journal} {Physical review letters}\ }\textbf {\bibinfo {volume}
  {118}},\ \bibinfo {pages} {166801} (\bibinfo {year} {2017})}\BibitemShut
  {NoStop}%
\bibitem [{\citenamefont {Garc{\' i}a-Mata}\ \emph {et~al.}(2020)\citenamefont
  {Garc{\' i}a-Mata}, \citenamefont {Martin}, \citenamefont {Dubertrand},
  \citenamefont {Giraud}, \citenamefont {Georgeot},\ and\ \citenamefont
  {Lemari{\'e}}}]{garc2019}%
  \BibitemOpen
  \bibfield  {author} {\bibinfo {author} {\bibfnamefont {I.}~\bibnamefont
  {Garc{\' i}a-Mata}}, \bibinfo {author} {\bibfnamefont {J.}~\bibnamefont
  {Martin}}, \bibinfo {author} {\bibfnamefont {R.}~\bibnamefont {Dubertrand}},
  \bibinfo {author} {\bibfnamefont {O.}~\bibnamefont {Giraud}}, \bibinfo
  {author} {\bibfnamefont {B.}~\bibnamefont {Georgeot}},\ and\ \bibinfo
  {author} {\bibfnamefont {G.}~\bibnamefont {Lemari{\'e}}},\ }\bibfield
  {title} {\bibinfo {title} {Two critical localization lengths in the anderson
  transition on random graphs},\ }\href@noop {} {\bibfield  {journal} {\bibinfo
   {journal} {Physical Review Research}\ }\textbf {\bibinfo {volume} {2}},\
  \bibinfo {pages} {012020} (\bibinfo {year} {2020})}\BibitemShut {NoStop}%
\bibitem [{\citenamefont {Biroli}\ and\ \citenamefont
  {Tarzia}(2018)}]{biroli2018delocalization}%
  \BibitemOpen
  \bibfield  {author} {\bibinfo {author} {\bibfnamefont {G.}~\bibnamefont
  {Biroli}}\ and\ \bibinfo {author} {\bibfnamefont {M.}~\bibnamefont
  {Tarzia}},\ }\bibfield  {title} {\bibinfo {title} {Delocalization and
  ergodicity of the anderson model on bethe lattices},\ }\href@noop {}
  {\bibfield  {journal} {\bibinfo  {journal} {arXiv preprint arXiv:1810.07545}\
  } (\bibinfo {year} {2018})}\BibitemShut {NoStop}%
\bibitem [{\citenamefont {Biroli}\ \emph {et~al.}(2022)\citenamefont {Biroli},
  \citenamefont {Hartmann},\ and\ \citenamefont {Tarzia}}]{Biroli2022}%
  \BibitemOpen
  \bibfield  {author} {\bibinfo {author} {\bibfnamefont {G.}~\bibnamefont
  {Biroli}}, \bibinfo {author} {\bibfnamefont {A.~K.}\ \bibnamefont
  {Hartmann}},\ and\ \bibinfo {author} {\bibfnamefont {M.}~\bibnamefont
  {Tarzia}},\ }\bibfield  {title} {\bibinfo {title} {Critical behavior of the
  anderson model on the bethe lattice via a large-deviation approach},\ }\href
  {https://doi.org/10.1103/PhysRevB.105.094202} {\bibfield  {journal} {\bibinfo
   {journal} {Phys. Rev. B}\ }\textbf {\bibinfo {volume} {105}},\ \bibinfo
  {pages} {094202} (\bibinfo {year} {2022})}\BibitemShut {NoStop}%
\bibitem [{\citenamefont {Mirlin}\ and\ \citenamefont
  {Fyodorov}(1994)}]{mirlin1994distribution}%
  \BibitemOpen
  \bibfield  {author} {\bibinfo {author} {\bibfnamefont {A.~D.}\ \bibnamefont
  {Mirlin}}\ and\ \bibinfo {author} {\bibfnamefont {Y.~V.}\ \bibnamefont
  {Fyodorov}},\ }\bibfield  {title} {\bibinfo {title} {Distribution of local
  densities of states, order parameter function, and critical behavior near the
  anderson transition},\ }\href@noop {} {\bibfield  {journal} {\bibinfo
  {journal} {Physical review letters}\ }\textbf {\bibinfo {volume} {72}},\
  \bibinfo {pages} {526} (\bibinfo {year} {1994})}\BibitemShut {NoStop}%
\bibitem [{\citenamefont {Mirlin}\ and\ \citenamefont
  {Fyodorov}(1991)}]{Mirlin_Fyodorov_1991}%
  \BibitemOpen
  \bibfield  {author} {\bibinfo {author} {\bibfnamefont {A.~D.}\ \bibnamefont
  {Mirlin}}\ and\ \bibinfo {author} {\bibfnamefont {Y.~V.}\ \bibnamefont
  {Fyodorov}},\ }\bibfield  {title} {\bibinfo {title} {Universality of level
  correlation function of sparse random matrices},\ }\href
  {https://doi.org/10.1088/0305-4470/24/10/016} {\bibfield  {journal} {\bibinfo
   {journal} {Journal of Physics A: Mathematical and General}\ }\textbf
  {\bibinfo {volume} {24}},\ \bibinfo {pages} {2273} (\bibinfo {year}
  {1991})}\BibitemShut {NoStop}%
\bibitem [{\citenamefont {Tikhonov}\ and\ \citenamefont
  {Mirlin}(2021)}]{Tikhonov_Mirlin_2021}%
  \BibitemOpen
  \bibfield  {author} {\bibinfo {author} {\bibfnamefont {K.}~\bibnamefont
  {Tikhonov}}\ and\ \bibinfo {author} {\bibfnamefont {A.}~\bibnamefont
  {Mirlin}},\ }\bibfield  {title} {\bibinfo {title} {From {Anderson}
  localization on random regular graphs to many-body localization},\ }\href
  {https://doi.org/10.1016/j.aop.2021.168525} {\bibfield  {journal} {\bibinfo
  {journal} {Annals of Physics}\ }\textbf {\bibinfo {volume} {435}},\ \bibinfo
  {pages} {168525} (\bibinfo {year} {2021})}\BibitemShut {NoStop}%
\bibitem [{\citenamefont {Tikhonov}\ and\ \citenamefont
  {Mirlin}(2018)}]{tikhonov2018}%
  \BibitemOpen
  \bibfield  {author} {\bibinfo {author} {\bibfnamefont {K.~S.}\ \bibnamefont
  {Tikhonov}}\ and\ \bibinfo {author} {\bibfnamefont {A.~D.}\ \bibnamefont
  {Mirlin}},\ }\bibfield  {title} {\bibinfo {title} {Many-body localization
  transition with power-law interactions: Statistics of eigenstates},\ }\href
  {https://doi.org/10.1103/PhysRevB.97.214205} {\bibfield  {journal} {\bibinfo
  {journal} {Phys. Rev. B}\ }\textbf {\bibinfo {volume} {97}},\ \bibinfo
  {pages} {214205} (\bibinfo {year} {2018})}\BibitemShut {NoStop}%
\bibitem [{\citenamefont {Efetov}(1985)}]{efetov1985anderson}%
  \BibitemOpen
  \bibfield  {author} {\bibinfo {author} {\bibfnamefont {K.}~\bibnamefont
  {Efetov}},\ }\bibfield  {title} {\bibinfo {title} {Anderson metal-insulator
  transition in a system of metal granules: Existence of a minimum metallic
  conductivity and a maximum dielectric constant},\ }\href@noop {} {\bibfield
  {journal} {\bibinfo  {journal} {Zh. Eksp. Teor. Fiz}\ }\textbf {\bibinfo
  {volume} {88}},\ \bibinfo {pages} {1032} (\bibinfo {year}
  {1985})}\BibitemShut {NoStop}%
\bibitem [{\citenamefont {Zirnbauer}(1986)}]{Zirnbauer1986}%
  \BibitemOpen
  \bibfield  {author} {\bibinfo {author} {\bibfnamefont {M.~R.}\ \bibnamefont
  {Zirnbauer}},\ }\bibfield  {title} {\bibinfo {title} {Localization transition
  on the bethe lattice},\ }\href {https://doi.org/10.1103/PhysRevB.34.6394}
  {\bibfield  {journal} {\bibinfo  {journal} {Phys. Rev. B}\ }\textbf {\bibinfo
  {volume} {34}},\ \bibinfo {pages} {6394} (\bibinfo {year}
  {1986})}\BibitemShut {NoStop}%
\bibitem [{\citenamefont {Biroli}\ and\ \citenamefont
  {Tarzia}(2017)}]{Biroli2017}%
  \BibitemOpen
  \bibfield  {author} {\bibinfo {author} {\bibfnamefont {G.}~\bibnamefont
  {Biroli}}\ and\ \bibinfo {author} {\bibfnamefont {M.}~\bibnamefont
  {Tarzia}},\ }\bibfield  {title} {\bibinfo {title} {Delocalized glassy
  dynamics and many-body localization},\ }\href
  {https://doi.org/10.1103/PhysRevB.96.201114} {\bibfield  {journal} {\bibinfo
  {journal} {Phys. Rev. B}\ }\textbf {\bibinfo {volume} {96}},\ \bibinfo
  {pages} {201114(R)} (\bibinfo {year} {2017})}\BibitemShut {NoStop}%
\bibitem [{\citenamefont {Edwards}\ and\ \citenamefont
  {Thouless}(1972)}]{Edwards1972}%
  \BibitemOpen
  \bibfield  {author} {\bibinfo {author} {\bibfnamefont {J.}~\bibnamefont
  {Edwards}}\ and\ \bibinfo {author} {\bibfnamefont {D.}~\bibnamefont
  {Thouless}},\ }\bibfield  {title} {\bibinfo {title} {Numerical studies of
  localization in disordered systems},\ }\href@noop {} {\bibfield  {journal}
  {\bibinfo  {journal} {Journal of Physics C: Solid State Physics}\ }\textbf
  {\bibinfo {volume} {5}},\ \bibinfo {pages} {807} (\bibinfo {year}
  {1972})}\BibitemShut {NoStop}%
\bibitem [{\citenamefont {Rodriguez}\ \emph {et~al.}(2010)\citenamefont
  {Rodriguez}, \citenamefont {Vasquez}, \citenamefont {Slevin},\ and\
  \citenamefont {R{\"o}mer}}]{rodriguez2010critical}%
  \BibitemOpen
  \bibfield  {author} {\bibinfo {author} {\bibfnamefont {A.}~\bibnamefont
  {Rodriguez}}, \bibinfo {author} {\bibfnamefont {L.~J.}\ \bibnamefont
  {Vasquez}}, \bibinfo {author} {\bibfnamefont {K.}~\bibnamefont {Slevin}},\
  and\ \bibinfo {author} {\bibfnamefont {R.~A.}\ \bibnamefont {R{\"o}mer}},\
  }\bibfield  {title} {\bibinfo {title} {Critical parameters from a generalized
  multifractal analysis at the anderson transition},\ }\href@noop {} {\bibfield
   {journal} {\bibinfo  {journal} {Physical review letters}\ }\textbf {\bibinfo
  {volume} {105}},\ \bibinfo {pages} {046403} (\bibinfo {year}
  {2010})}\BibitemShut {NoStop}%
\bibitem [{\citenamefont {Ortu{\~n}o}\ \emph {et~al.}(2009)\citenamefont
  {Ortu{\~n}o}, \citenamefont {Somoza},\ and\ \citenamefont
  {Chalker}}]{ortuno2009random}%
  \BibitemOpen
  \bibfield  {author} {\bibinfo {author} {\bibfnamefont {M.}~\bibnamefont
  {Ortu{\~n}o}}, \bibinfo {author} {\bibfnamefont {A.}~\bibnamefont {Somoza}},\
  and\ \bibinfo {author} {\bibfnamefont {J.}~\bibnamefont {Chalker}},\
  }\bibfield  {title} {\bibinfo {title} {Random walks and anderson localization
  in a three-dimensional class c network model},\ }\href@noop {} {\bibfield
  {journal} {\bibinfo  {journal} {Physical review letters}\ }\textbf {\bibinfo
  {volume} {102}},\ \bibinfo {pages} {070603} (\bibinfo {year}
  {2009})}\BibitemShut {NoStop}%
\bibitem [{\citenamefont {Hernandez}\ \emph {et~al.}(2005)\citenamefont
  {Hernandez}, \citenamefont {Roman},\ and\ \citenamefont
  {Vidal}}]{Hernandez:2005:SSF}%
  \BibitemOpen
  \bibfield  {author} {\bibinfo {author} {\bibfnamefont {V.}~\bibnamefont
  {Hernandez}}, \bibinfo {author} {\bibfnamefont {J.~E.}\ \bibnamefont
  {Roman}},\ and\ \bibinfo {author} {\bibfnamefont {V.}~\bibnamefont {Vidal}},\
  }\bibfield  {title} {\bibinfo {title} {{SLEPc}: A scalable and flexible
  toolkit for the solution of eigenvalue problems},\ }\href@noop {} {\bibfield
  {journal} {\bibinfo  {journal} {{ACM} Trans. Math. Software}\ }\textbf
  {\bibinfo {volume} {31}},\ \bibinfo {pages} {351} (\bibinfo {year}
  {2005})}\BibitemShut {NoStop}%
\bibitem [{\citenamefont {Roman}\ and\ \citenamefont
  {Pino}(tion)}]{Roman2024_preparation}%
  \BibitemOpen
  \bibfield  {author} {\bibinfo {author} {\bibfnamefont {J.~E.}\ \bibnamefont
  {Roman}}\ and\ \bibinfo {author} {\bibfnamefont {M.}~\bibnamefont {Pino}},\
  }\bibfield  {title} {\bibinfo {title} {Large diagonalizations via polynomial
  filters in {SLEP}c: application to eigenvectors of the {A}nderson model in a
  random-regular graph}} (\bibinfo {year} {in preparation})\BibitemShut
  {NoStop}%
\bibitem [{\citenamefont {Pino}(2020)}]{pino2020scaling}%
  \BibitemOpen
  \bibfield  {author} {\bibinfo {author} {\bibfnamefont {M.}~\bibnamefont
  {Pino}},\ }\bibfield  {title} {\bibinfo {title} {Scaling up the {Anderson}
  transition in random-regular graphs},\ }\href@noop {} {\bibfield  {journal}
  {\bibinfo  {journal} {Physical Review Research}\ }\textbf {\bibinfo {volume}
  {2}},\ \bibinfo {pages} {042031} (\bibinfo {year} {2020})}\BibitemShut
  {NoStop}%
\bibitem [{\citenamefont {Anderson}(1958)}]{An1958}%
  \BibitemOpen
  \bibfield  {author} {\bibinfo {author} {\bibfnamefont {P.~W.}\ \bibnamefont
  {Anderson}},\ }\bibfield  {title} {\bibinfo {title} {Absence of diffusion in
  certain random lattices},\ }\href {https://doi.org/10.1103/PhysRev.109.1492}
  {\bibfield  {journal} {\bibinfo  {journal} {Phys. Rev.}\ }\textbf {\bibinfo
  {volume} {109}},\ \bibinfo {pages} {1492} (\bibinfo {year}
  {1958})}\BibitemShut {NoStop}%
\bibitem [{\citenamefont {Abrahams}\ \emph {et~al.}(1979)\citenamefont
  {Abrahams}, \citenamefont {Anderson}, \citenamefont {Licciardello},\ and\
  \citenamefont {Ramakrishnan}}]{Ab1979}%
  \BibitemOpen
  \bibfield  {author} {\bibinfo {author} {\bibfnamefont {E.}~\bibnamefont
  {Abrahams}}, \bibinfo {author} {\bibfnamefont {P.~W.}\ \bibnamefont
  {Anderson}}, \bibinfo {author} {\bibfnamefont {D.~C.}\ \bibnamefont
  {Licciardello}},\ and\ \bibinfo {author} {\bibfnamefont {T.~V.}\ \bibnamefont
  {Ramakrishnan}},\ }\bibfield  {title} {\bibinfo {title} {Scaling theory of
  localization: Absence of quantum diffusion in two dimensions},\ }\href
  {https://doi.org/10.1103/PhysRevLett.42.673} {\bibfield  {journal} {\bibinfo
  {journal} {Phys. Rev. Lett.}\ }\textbf {\bibinfo {volume} {42}},\ \bibinfo
  {pages} {673} (\bibinfo {year} {1979})}\BibitemShut {NoStop}%
\bibitem [{\citenamefont {Hagberg}\ \emph {et~al.}(2008)\citenamefont
  {Hagberg}, \citenamefont {Swart},\ and\ \citenamefont
  {S~Chult}}]{pythonnetwokx}%
  \BibitemOpen
  \bibfield  {author} {\bibinfo {author} {\bibfnamefont {A.}~\bibnamefont
  {Hagberg}}, \bibinfo {author} {\bibfnamefont {P.}~\bibnamefont {Swart}},\
  and\ \bibinfo {author} {\bibfnamefont {D.}~\bibnamefont {S~Chult}},\
  }\href@noop {} {\emph {\bibinfo {title} {Exploring network structure,
  dynamics, and function using NetworkX}}},\ \bibinfo {type} {Tech. Rep.}\
  (\bibinfo  {institution} {Los Alamos National Lab.(LANL), Los Alamos, NM
  (United States)},\ \bibinfo {year} {2008})\BibitemShut {NoStop}%
\bibitem [{\citenamefont {Sierant}\ \emph {et~al.}(2023)\citenamefont
  {Sierant}, \citenamefont {Lewenstein},\ and\ \citenamefont
  {Scardicchio}}]{sierant2023universality}%
  \BibitemOpen
  \bibfield  {author} {\bibinfo {author} {\bibfnamefont {P.}~\bibnamefont
  {Sierant}}, \bibinfo {author} {\bibfnamefont {M.}~\bibnamefont
  {Lewenstein}},\ and\ \bibinfo {author} {\bibfnamefont {A.}~\bibnamefont
  {Scardicchio}},\ }\bibfield  {title} {\bibinfo {title} {Universality in
  anderson localization on random graphs with varying connectivity},\
  }\href@noop {} {\bibfield  {journal} {\bibinfo  {journal} {SciPost Physics}\
  }\textbf {\bibinfo {volume} {15}},\ \bibinfo {pages} {045} (\bibinfo {year}
  {2023})}\BibitemShut {NoStop}%
\bibitem [{\citenamefont {Halsey}\ \emph {et~al.}(1986)\citenamefont {Halsey},
  \citenamefont {Jensen}, \citenamefont {Kadanoff}, \citenamefont {Procaccia},\
  and\ \citenamefont {Shraiman}}]{Kadanoff1986}%
  \BibitemOpen
  \bibfield  {author} {\bibinfo {author} {\bibfnamefont {T.~C.}\ \bibnamefont
  {Halsey}}, \bibinfo {author} {\bibfnamefont {M.~H.}\ \bibnamefont {Jensen}},
  \bibinfo {author} {\bibfnamefont {L.~P.}\ \bibnamefont {Kadanoff}}, \bibinfo
  {author} {\bibfnamefont {I.}~\bibnamefont {Procaccia}},\ and\ \bibinfo
  {author} {\bibfnamefont {B.~I.}\ \bibnamefont {Shraiman}},\ }\bibfield
  {title} {\bibinfo {title} {Fractal measures and their singularities: The
  characterization of strange sets},\ }\href
  {https://doi.org/10.1103/PhysRevA.33.1141} {\bibfield  {journal} {\bibinfo
  {journal} {Phys. Rev. A}\ }\textbf {\bibinfo {volume} {33}},\ \bibinfo
  {pages} {1141} (\bibinfo {year} {1986})}\BibitemShut {NoStop}%
\bibitem [{\citenamefont {Rodriguez}\ \emph {et~al.}(2011)\citenamefont
  {Rodriguez}, \citenamefont {Vasquez}, \citenamefont {Slevin},\ and\
  \citenamefont {R\"omer}}]{rodriguez2011}%
  \BibitemOpen
  \bibfield  {author} {\bibinfo {author} {\bibfnamefont {A.}~\bibnamefont
  {Rodriguez}}, \bibinfo {author} {\bibfnamefont {L.~J.}\ \bibnamefont
  {Vasquez}}, \bibinfo {author} {\bibfnamefont {K.}~\bibnamefont {Slevin}},\
  and\ \bibinfo {author} {\bibfnamefont {R.~A.}\ \bibnamefont {R\"omer}},\
  }\bibfield  {title} {\bibinfo {title} {Multifractal finite-size scaling and
  universality at the anderson transition},\ }\href
  {https://doi.org/10.1103/PhysRevB.84.134209} {\bibfield  {journal} {\bibinfo
  {journal} {Phys. Rev. B}\ }\textbf {\bibinfo {volume} {84}},\ \bibinfo
  {pages} {134209} (\bibinfo {year} {2011})}\BibitemShut {NoStop}%
\bibitem [{\citenamefont {Pausch}\ \emph {et~al.}(2021)\citenamefont {Pausch},
  \citenamefont {Carnio}, \citenamefont {Rodr{\'\i}guez},\ and\ \citenamefont
  {Buchleitner}}]{pausch2021chaos}%
  \BibitemOpen
  \bibfield  {author} {\bibinfo {author} {\bibfnamefont {L.}~\bibnamefont
  {Pausch}}, \bibinfo {author} {\bibfnamefont {E.~G.}\ \bibnamefont {Carnio}},
  \bibinfo {author} {\bibfnamefont {A.}~\bibnamefont {Rodr{\'\i}guez}},\ and\
  \bibinfo {author} {\bibfnamefont {A.}~\bibnamefont {Buchleitner}},\
  }\bibfield  {title} {\bibinfo {title} {Chaos and ergodicity across the energy
  spectrum of interacting bosons},\ }\href@noop {} {\bibfield  {journal}
  {\bibinfo  {journal} {Physical Review Letters}\ }\textbf {\bibinfo {volume}
  {126}},\ \bibinfo {pages} {150601} (\bibinfo {year} {2021})}\BibitemShut
  {NoStop}%
\bibitem [{\citenamefont {Cardy}(1996)}]{cardy1996scaling}%
  \BibitemOpen
  \bibfield  {author} {\bibinfo {author} {\bibfnamefont {J.}~\bibnamefont
  {Cardy}},\ }\href@noop {} {\emph {\bibinfo {title} {Scaling and
  renormalization in statistical physics}}},\ Vol.~\bibinfo {volume} {5}\
  (\bibinfo  {publisher} {Cambridge university press},\ \bibinfo {year}
  {1996})\BibitemShut {NoStop}%
\bibitem [{\citenamefont {Baroni}\ \emph {et~al.}(2023)\citenamefont {Baroni},
  \citenamefont {Lorenzana}, \citenamefont {Rizzo},\ and\ \citenamefont
  {Tarzia}}]{baroni2023corrections}%
  \BibitemOpen
  \bibfield  {author} {\bibinfo {author} {\bibfnamefont {M.}~\bibnamefont
  {Baroni}}, \bibinfo {author} {\bibfnamefont {G.~G.}\ \bibnamefont
  {Lorenzana}}, \bibinfo {author} {\bibfnamefont {T.}~\bibnamefont {Rizzo}},\
  and\ \bibinfo {author} {\bibfnamefont {M.}~\bibnamefont {Tarzia}},\
  }\bibfield  {title} {\bibinfo {title} {Corrections to the bethe lattice
  solution of anderson localization},\ }\href@noop {} {\bibfield  {journal}
  {\bibinfo  {journal} {arXiv preprint arXiv:2304.10365}\ } (\bibinfo {year}
  {2023})}\BibitemShut {NoStop}%
\bibitem [{\citenamefont {Abou-Chacra}\ \emph {et~al.}(1973)\citenamefont
  {Abou-Chacra}, \citenamefont {Thouless},\ and\ \citenamefont
  {Anderson}}]{abou1973selfconsistent}%
  \BibitemOpen
  \bibfield  {author} {\bibinfo {author} {\bibfnamefont {R.}~\bibnamefont
  {Abou-Chacra}}, \bibinfo {author} {\bibfnamefont {D.}~\bibnamefont
  {Thouless}},\ and\ \bibinfo {author} {\bibfnamefont {P.}~\bibnamefont
  {Anderson}},\ }\bibfield  {title} {\bibinfo {title} {A selfconsistent theory
  of localization},\ }\href@noop {} {\bibfield  {journal} {\bibinfo  {journal}
  {Journal of Physics C: Solid State Physics}\ }\textbf {\bibinfo {volume}
  {6}},\ \bibinfo {pages} {1734} (\bibinfo {year} {1973})}\BibitemShut
  {NoStop}%
\bibitem [{\citenamefont {Andrae}\ \emph {et~al.}(2010)\citenamefont {Andrae},
  \citenamefont {Schulze-Hartung},\ and\ \citenamefont
  {Melchior}}]{andrae2010}%
  \BibitemOpen
  \bibfield  {author} {\bibinfo {author} {\bibfnamefont {R.}~\bibnamefont
  {Andrae}}, \bibinfo {author} {\bibfnamefont {T.}~\bibnamefont
  {Schulze-Hartung}},\ and\ \bibinfo {author} {\bibfnamefont {P.}~\bibnamefont
  {Melchior}},\ }\bibfield  {title} {\bibinfo {title} {Dos and don'ts of
  reduced chi-squared},\ }\href@noop {} {\bibfield  {journal} {\bibinfo
  {journal} {arXiv preprint arXiv:1012.3754}\ } (\bibinfo {year}
  {2010})}\BibitemShut {NoStop}%
\end{thebibliography}%

\appendix

\section{Correlated volume from the imaginary part of the Green's function}\label{sec:app2}
\begin{figure*}[t!]
\includegraphics[width=0.8\textwidth]{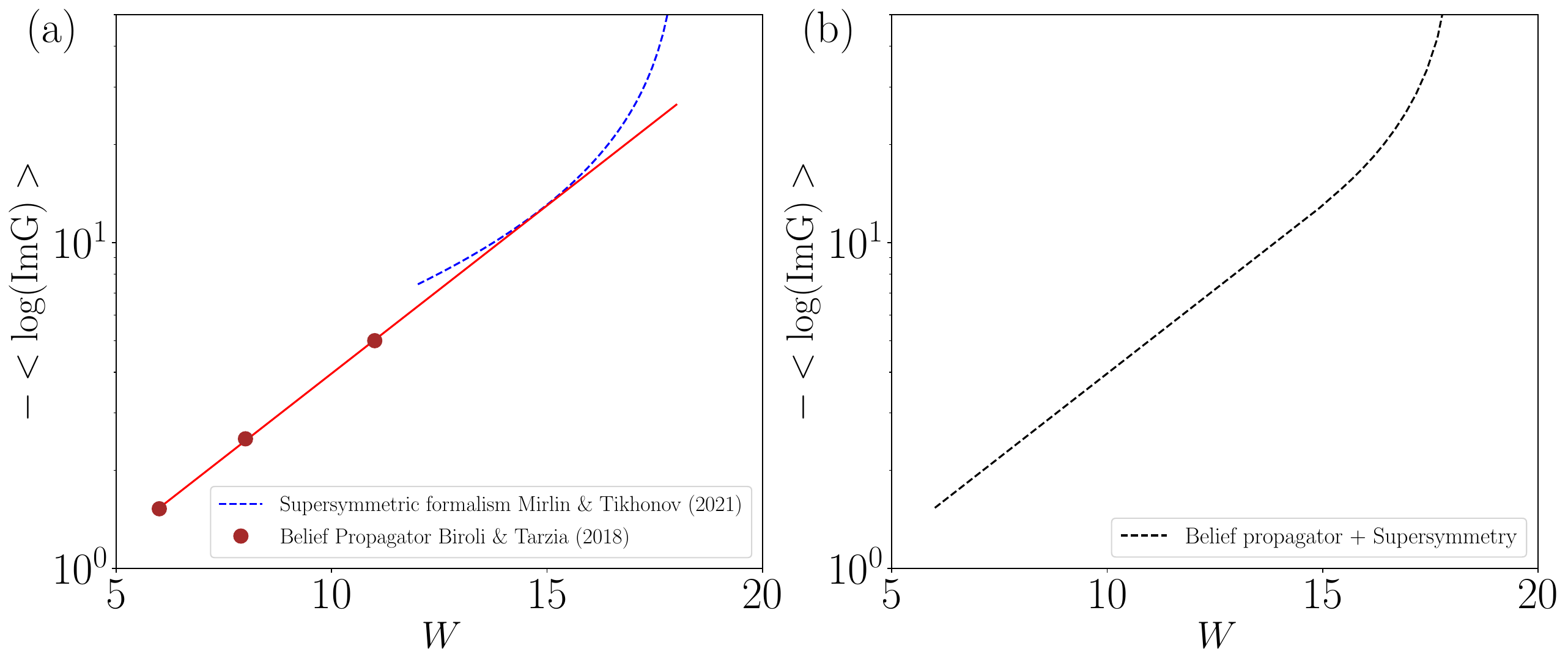}
\caption{\label{Fig:S1} (a) Mean-field results for minus the logarithm of the typical value for the imaginary part of Green's function in a random-regular graph with branching number $k=2.$ The bullets are the results from population dynamics contained in Ref. \cite{biroli2018delocalization}. The red solid line is an extrapolation of these points to a straight line having a slope of approximately $0.24$.  The blue solid line is obtained via the super-symmetric formalism Eq.~(23) of Ref.\ \cite{Tikhonov_Mirlin_2021}. (b) Interpolation of the results for the population dynamics of Ref. \cite{biroli2018delocalization} in the range $W\in [6,15]$ and the super-symmetric formalism in $W\in [15,18.2].$ }
\end{figure*}

We explain how we have interpolated previous results obtained by Tarzia \& Biroli via Population Dynamics\ \cite{biroli2018delocalization} and the ones by Mirlin \& Tikhonov using the super-symmetric formalism\ \cite{Tikhonov_Mirlin_2021}. From the last reference Eq.~(23), the correlated volume is computed up  to second-order near the Anderson phase transition $W_c=18.17$ from the typical value of the imaginary part of the diagonal Greens function:
\begin{align}
  {\rm<Im G^{-}>_{typ}} =e^{-\left[a_1\left(W_c-W\right)^{1/2}+a_2\left(W_c-W\right)^{3/2}\right]^{-1}  }\label{eq:S1}
\end{align}
with $a_1=0.0313$ and $a_2=0.00369.$ The ergodic volume can be extracted as $\log(N_e) = -\log({\rm<Im G^->_{typ}}),$ which is related to the typical value of the local density of states $\rho_{E=0}(r)$ at the middle of the spectrum, as $\rho_E(r)=\mp {\rm Im{G^{(\pm)}}(r,r)}/\pi$  The same quantity can be obtained via population dynamics algorithms which, roughly speaking, find numerically a distribution probability that solves the self-consistent equations in the Bethe lattice for the Green's function\ \cite{abou1973selfconsistent}. Doing so, Biroli \& Tarzia obtained the diagonal part of $G^-$ and thus its typical value for which we have represented a few points (red scatters) in Fig.\ \ref{Fig:S1} (a).  We have fitted those points to a linear function in double log coordinates in the x-axis. The result is represented as a red solid line. In the same panel, we show the super-symmetric law Eq. \ref{eq:S1}~\cite{Tikhonov_Mirlin_2021} as a blue solid line from $W=12$ to $W=18.$

The result from Eq.\ \ref{eq:S1}, dashed blue line in Fig.\ \ref{Fig:S1} (a), have a different tendency for disorder around $W\approx 12$ than the linear tendency of the data from Biroli and Tarzia, solid red line in Fig.\ \ref{Fig:S1} (a). This is not a surprise as the supersymmetric formula is valid only close enough to the transition, as it is a second-order expansion in  $(W-W_c)^{-1/2}.$ Thus, we take as valid the extrapolated line for the belief propagator data of Ref.\ \cite{biroli2018delocalization} at a small disorder, while the supersymmetric formula at a larger disorder. We take the point where the solid and dashed in Fig.\ \ref{Fig:S1} (a) coincide as separating the region of validity for each of the laws. We reconstruct the ``mean-field'' solution as the line appearing in Fig.\ \ref{Fig:S1}(b), which is the one shown in several of the plots of the main body of the paper. We have checked that results from population dynamics contained in Ref.\ \cite{Tikhonov_Mirlin_2021} are very well described by this reconstruction of the ``mean-field'' solution.

\section{Analysis of the finite-size corrections to fractal dimensions and multifractal spectrum maximum}\label{supp:fittings}

We explore different functional forms to fit effective fractal dimensions $\widetilde{D}_q =\log_N(I_q)/(q-1)$ and the exponential of the typical value of the wavefunctions amplitude $\widetilde{\alpha}_0=\av{\log_N(|\psi|^2)}.$ The $N$-dependence of those two quantities are captured by Eq.\ \ref{eq:corr} of the main text. However, we can only set a reduced number of terms in those expressions in order to avoid an overfit due to the limited number of data points available. We comment here on how many and which corrections should be included to obtain reliable thermodynamic extrapolations. We also provide additional information to the fits shown in Fig.\ \ref{fig:D1} of the main body of the text.

\begin{figure*}[t!]
\includegraphics[width=0.6\textwidth]{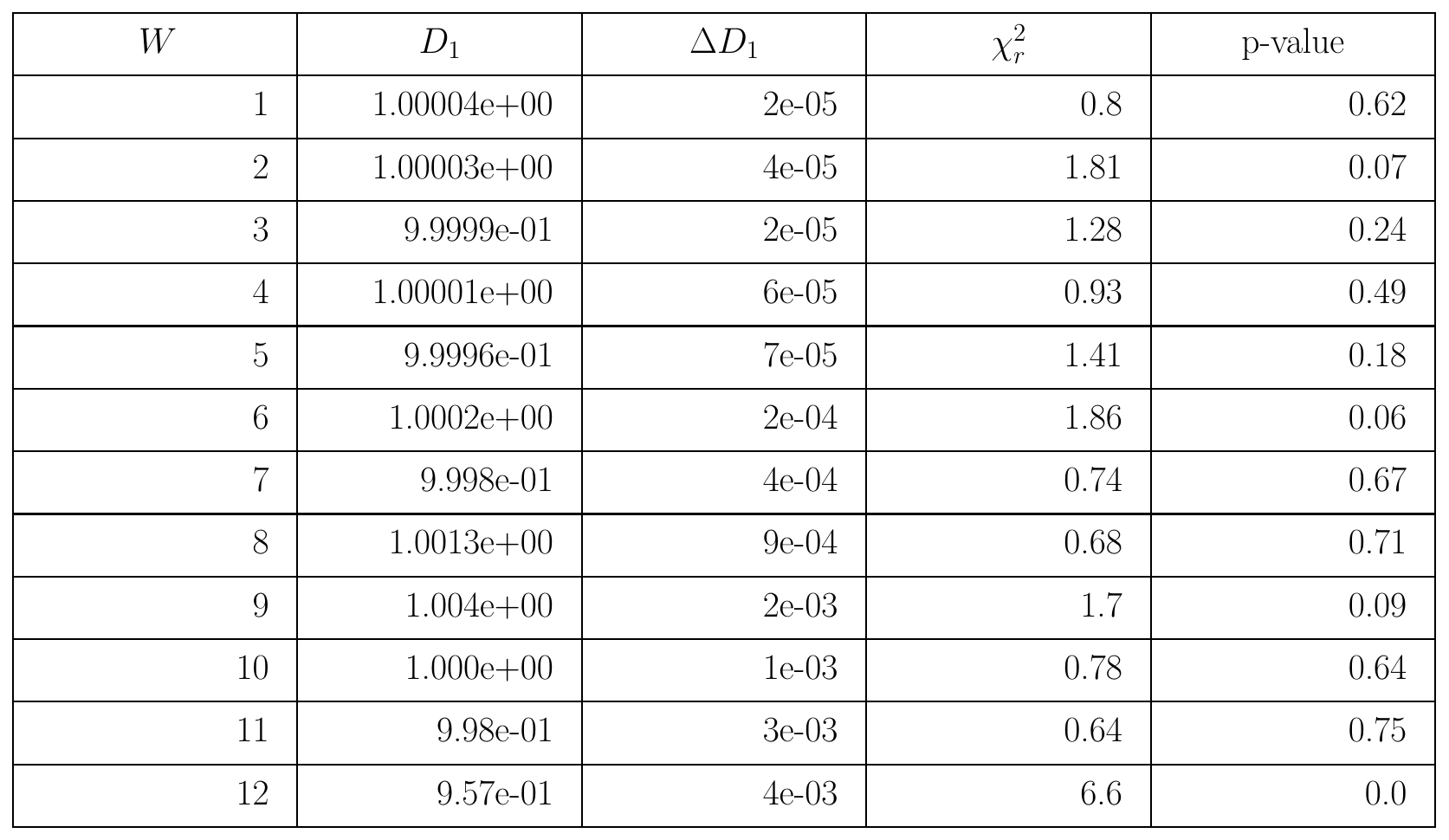}
\caption{\label{Fig:D1_details} Additional information for the fittings of the effective first fractal dimension $\widetilde{D}_1 =S/\log(N),$ with $S$ participation entropy, to a law $\sum_{j=0,k=0}^{r,s} \frac{a_{jk}(W)}{N^j[\log(N)]^k}$ with $r=s=1$ for sizes $N=10^3,\dots,4\times 10^6$  and several values of the disorder $W$.  The first column is the disorder, the second (third) are fractal dimensions (errors) extracted from the fitting parameters as $D_1=a_{00}$. The last two columns are the reduced-$\chi^2$ and the $p$-value of each fit.
}
\end{figure*}

We begin with the details of the fitting for $\widetilde{D}_1$ in Fig.\ \ref{fig:D1} of the main text. The data for this quantity was fitted to the law Eq.\ \ref{eq:corr} of the main text with $r=s=1$. In Fig.\ \ref{Fig:D1_details},  we show additional information regarding that fit.  As explained in the main text, the quality of the fitting is good up to $W=12$, where the smallness of the $p$-value implies that our law does not capture the size dependence of our data. Almost all extracted fractal dimensions with acceptable goodness of fit ($p$-values larger than $0.1$) are consistent with $D_1=1$. Thus, the data set for which the goodness of fit is acceptable gives meaningful parameters results, as we expect to have ergodicity $D_1=1$ deep enough in the metal. We are going to see in next paragraph that this is not the case when fitting with other corrections, different than $r=s=1$.

\begin{figure*}[t]
\includegraphics[width=0.6\textwidth]{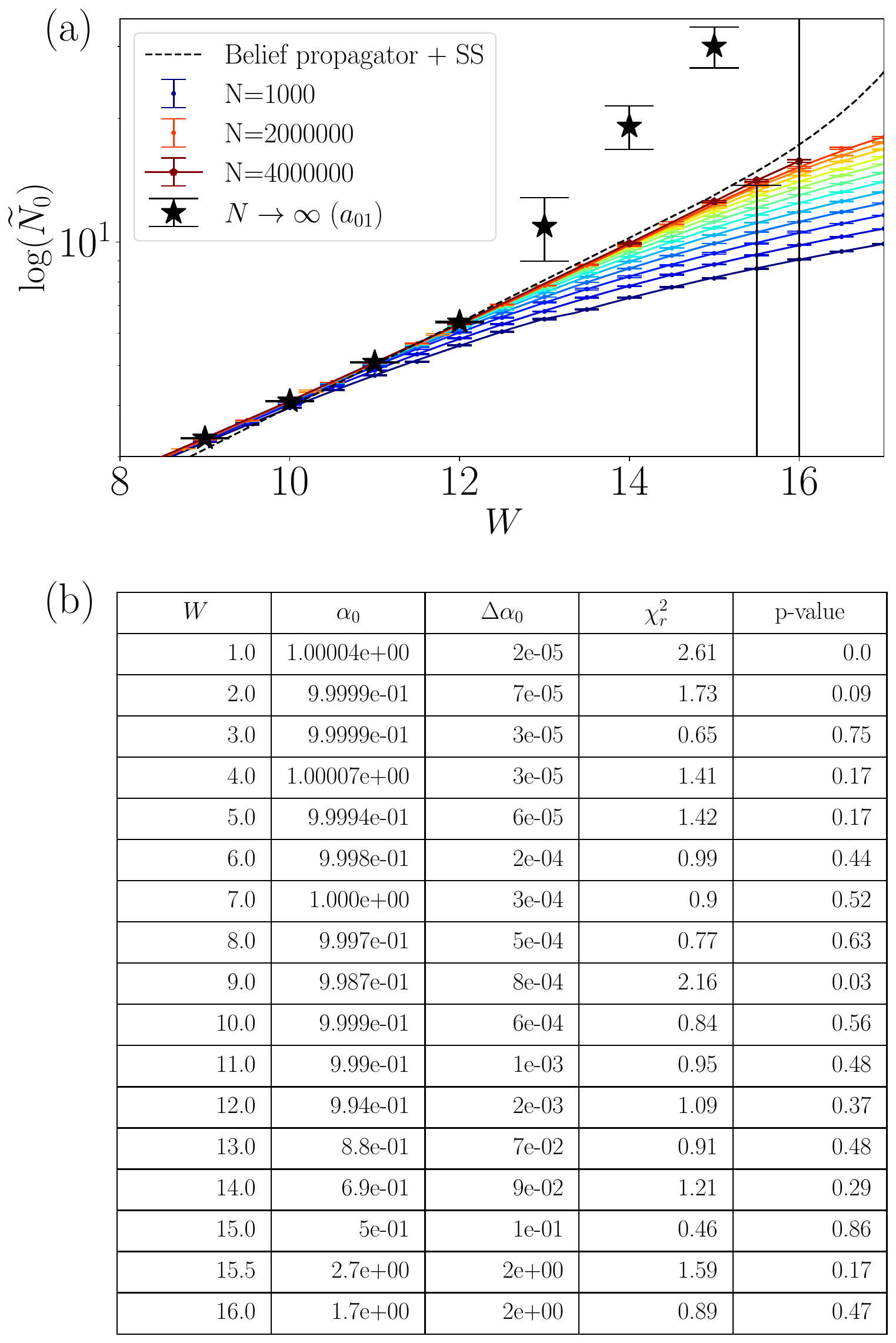}
\caption{\label{Fig:alpha_f1} (a) Logarithm of the effective correlated volume $\log(N_0)$ as a function of disorder for several system sizes. Dashed line is  the typical imaginary part of the self-energy obtained via belief propagation and super-symmetric formalism in Refs.\ \cite{biroli2018delocalization, Tikhonov_Mirlin_2021}. Stars are the value of $\log(\widetilde{N}_q)$ in the thermodynamic limit extracted from the coefficient $\log(N_0) = a^{(0)}_{01}$ when fitting $\widetilde{\alpha}_0=\sum_{j=0,k=0}^{1} a^{(0)}_{jk}(W)/(N^j[\log(N)]^k)$ for all the available sizes $N=10^6,\dots,4\times 10^4$ at each disorder value. The $r,s$ are chosen to minimize $|1-\chi_r^2|,$ where $\chi_r^2$ is the reduced chi-square statistic. That is $r=s=1$ up to $W=12,$ $r=1$ and $s=2$ from $12<W<15$ and $r=s=2$ for the largest disorder.  (b) Parameters of the fittings. The first column is the disorder, the second and third are the spectrum maximum with its errors and the last two columns are the reduced-$\chi^2$ and the $p$-value of each fit.
}
\end{figure*}

Now we comment on the fittings to extract the multifractal spectrum maximum $\alpha_0$ and its associated correlated volume $N_0.$  We have seen in the main text that $\widetilde{\alpha}=\av{\log_N(|\psi|^2)}$ shows small finite-size effects for the largest sizes up to disorder $W=15.$ Nevertheless, we have only been able to obtain acceptable fits of $\widetilde{\alpha}_0$ to the law in Eq.\ \ref{eq:corr} up to $W=12$ in Fig.\ \ref{fig:loglo_x1}  of the main text. This is in part due to the way we perform the fits with all available sizes and only use a small number of corrections in Eq.\ \ref{eq:corr}, indeed those with  $r=s=1.$ It is clear that a larger number of corrections should be needed when approaching the Anderson transition. Thus, our first procedure to extract thermodynamic limit extrapolation has been to increase the number of corrections. We did so by using values of $r,s$ that produce the fit with the reduced-$\chi^2$ statistic closest to one\ \cite{andrae2010}. Fig.\ \ref{Fig:alpha_f1} (a) shows the extrapolated value of the fits for the logarithm of the zero-correlated volume $\log(N_0)$. It produces values that are not expected at all on physical grounds. Besides  a too large correlated volume $\log(N_0)$, notice the log scale in the $y$-axis of Fig.\ \ref{Fig:alpha_f1} (a), it predicts the position of the multifractal spectrum $\alpha_0<1$ which is not possible, see\ \ref{Fig:alpha_f1} (b). We remark that no hint of a poor quality fitting can be inferred from the $p$-value or the $\chi_r^2$ statistics for the fittings that provided non-physical results, see\ \ref{Fig:alpha_f1} (b). This is different from results commented in the previous paragraph, where non-physical results could be mostly pointed out by bad quality fittings.

\begin{figure*}[t]
\includegraphics[width=0.8\textwidth]{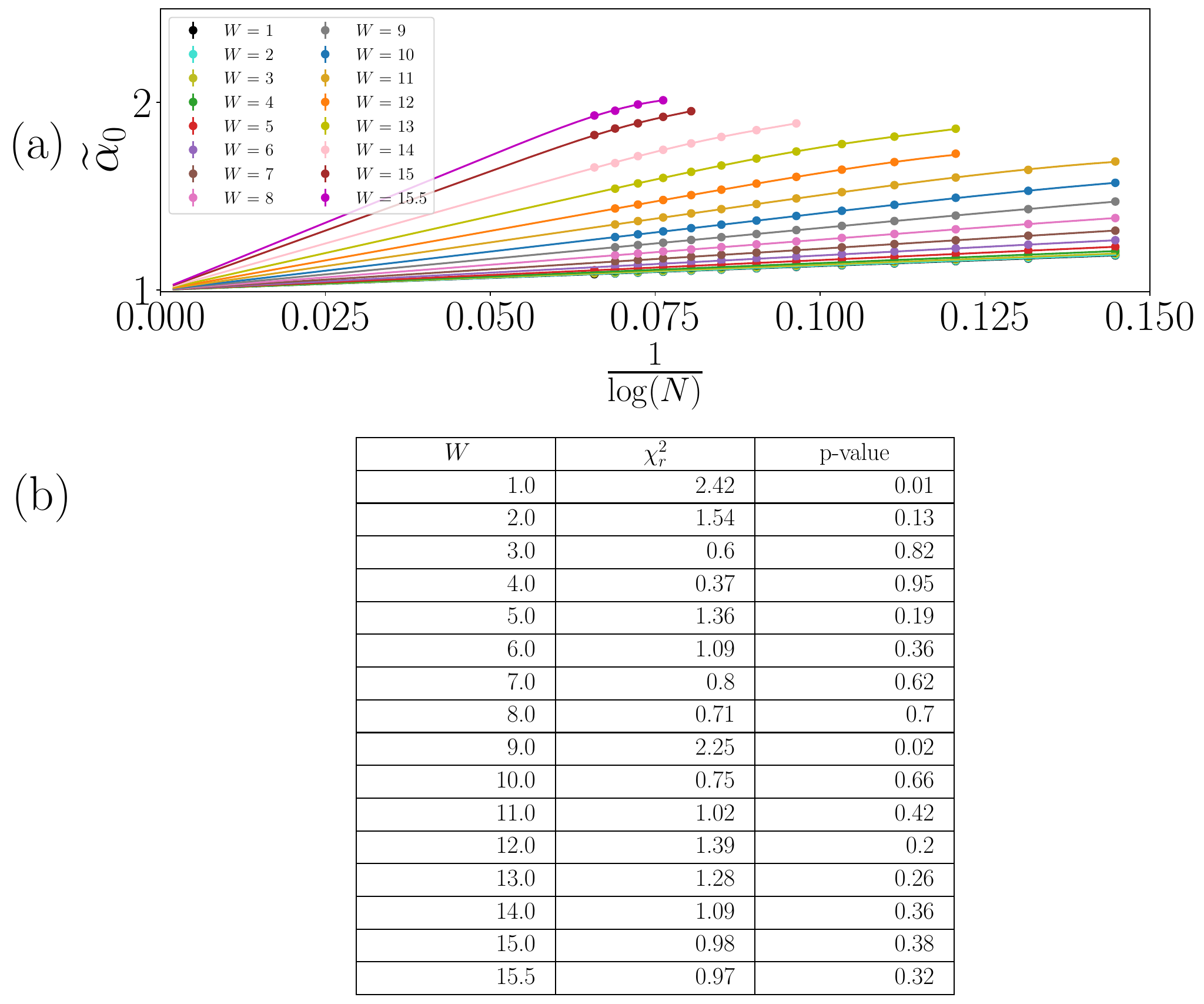}
\caption{\label{Fig:alpha_f2} (a) Logarithm of the effective location of the multifractal spectrum $\widetilde{\alpha}_0=\av{\log_N(|\psi|^2)}$ as a function of $1/\log(N),$ being $N$ the system size. Solid lines are a fit of the data to the law Eq.\ \ref{eq:corr} with the ergodic value fixed $a_{00}=1$ and corrections $r=s=1.$ The number of points that are fitted are those which produced the closest to one $\chi_r^2.$ (b) Parameters of the fits. The first column is the disorder, the second is the reduced-$\chi^2$ and the third is the $p$-value of each fit.
}
\end{figure*}

Similarly to the previous discussion, we have obtained good quality fittings whose parameters are non-physical, when fitting fractal dimensions $D_q$ with $q=1,2$  with $r,s>1$ in Eq.\ \ref{eq:corr}. This behaviour should be attributed to the slowness of the corrections given by a power law in $1/\log(N)$, and it is the reason why we prefer to fit the number of corrections and allow a smaller number of data points. As explained in the main text, we fitted numerical results for $\widetilde{\alpha}=\av{\log_N(|\psi|^2)}$ to  Eq.\ \ref{eq:corr} of the main text with fixed $r=s=1,$ but employing only the largest system sizes in order to obtain the $\chi_r^2$ closest to one. To help the fitting procedure, we have also fixed the thermodynamic value $\alpha_0=1$ in the fitting law. With this procedure we obtained the correlated volume $N_0$ displayed in Fig.\  \ref{fig:N0} of the main text. We provide additional information for those fittings in Fig.\ \ref{Fig:alpha_f2}, as the $p$-value or the $\chi_r^2$ statistics. Notice that the $\chi_r^2$ does not indicate an overfit for any of the disorder values, second column of panel (b).

\end{document}